\documentclass[reprint,NumberedRefs]{JASAnew}

\usepackage[nolist]{acronym}            % acronym expansion (no list printed)
\usepackage{siunitx}                    % \unit{\decibel}, \SI{...}{...}
\usepackage{multirow}                   % \multirow in tables
\usepackage{makecell}                   % \makecell in tables
\usepackage{booktabs}                   % \toprule, \midrule, \bottomrule

\graphicspath{{./}{plots/}}

\newif\ifshowchanges
\showchangesfalse % no blue color, original black
\DeclareRobustCommand{\chg}[1]{%
  \ifshowchanges{\color{blue}#1}\else #1\fi%
}

\newif\ifshowrev
\showrevfalse   % no highlights, all black
\DeclareRobustCommand{\rev}[1]{%
  \ifshowrev{\color{blue}#1}\else #1\fi%
}

\begin{document}

\title[JASA/Neural directional filtering]{Neural directional filtering using a compact microphone array}

\author{Weilong Huang}
\email{weilong.huang@audiolabs-erlangen.de}
\affiliation{International Audio Laboratories Erlangen (a joint institution of Fraunhofer IIS and Friedrich-Alexander-Universit\"{a}t Erlangen-N\"{u}rnberg, FAU, Germany), Erlangen, 91058, Germany}

\author{Srikanth Raj Chetupalli}
\thanks{Now at: Indian Institute of Technology Bombay, Mumbai, Maharashtra 400076, India.}
\affiliation{Fraunhofer Institute for Integrated Circuits IIS, Erlangen, 91058, Germany}

\author{Mhd Modar Halimeh}
\thanks{Now at: Starkey Hearing Technologies, Eden Prairie, Minnesota 55344, United States.}
\affiliation{Fraunhofer Institute for Integrated Circuits IIS, Erlangen, 91058, Germany}

% \author{Srikanth Raj Chetupalli}
% \affiliation{Department of Electrical Engineering, Indian Institute of Technology Bombay, Mumbai, Maharashtra 400076, India}

% \author{Mhd Modar Halimeh}
% \affiliation{Starkey Hearing Technologies, Eden Prairie, Minnesota 55344, United States}

\author{Oliver Thiergart}
\affiliation{Fraunhofer Institute for Integrated Circuits IIS, Erlangen, 91058, Germany}

\author{Emanu\"{e}l A. P. Habets}
\affiliation{International Audio Laboratories Erlangen (a joint institution of Fraunhofer IIS and Friedrich-Alexander-Universit\"{a}t Erlangen-N\"{u}rnberg, FAU, Germany), Erlangen, 91058, Germany}

\preprint{Huang \emph{et al.}, JASA}

\date{\today}

\begin{abstract}
Beamforming with desired directivity patterns using compact microphone arrays is essential in many audio applications. Directivity patterns achievable using traditional beamformers depend on the number of microphones and the array aperture. Generally, their effectiveness degrades for compact arrays. Neural directional filtering (NDF) overcomes these limitations by leveraging deep neural networks to enable sound capture with a predefined directivity pattern. The NDF computes a single-channel complex mask from the microphone array signals, which is then applied to a reference microphone to produce an output that approximates the output of a virtual directional microphone with the desired directivity pattern. This paper presents a comprehensive study of NDF, introducing training strategies and data-dependent metrics for evaluating the directivity pattern and the directivity factor. It is shown that the NDF: i) achieves a frequency-invariant directivity pattern even above the spatial aliasing frequency, ii) can approximate diverse and higher-order patterns, iii) can steer the pattern in different directions, and iv) generalizes to unseen conditions. Lastly, experimental comparisons demonstrate superior performance over conventional beamforming and parametric approaches.
\end{abstract}

%% pacs numbers not used

\maketitle

%% Acronym definitions are placed after \maketitle so that they are
%% available throughout the body but do not conflict with revtex's
%% pre-processing of the title/abstract block.
\begin{acronym}[DNN]
	\acro{DNN}[DNN]{deep neural network}
\end{acronym}
\begin{acronym}[iSTFT]
	\acro{sdri}[$\Delta$SDR]{improvement in \ac{SDR} over the unprocessed signal}
	\acro{DMA}[DMA]{differential microphone array}
	\acro{DNN}[DNN]{deep neural network}
	\acro{DOA}[DOA]{direction-of-arrival}
	\acrodefplural{DOA}[DOAs]{directions-of-arrival}
	\acro{iSTFT}[iSTFT]{inverse short-time Fourier transform}
	\acro{CDMA}[CDMA]{circular \ac{DMA}}
	\acro{LDMA}[LDMA]{linear \ac{DMA}}
	\acro{LS}{least-squares}
	\acro{LSTM}[LSTM]{long short-term memory}
	\acro{BiLSTM}[BiLSTM]{bidirectional LSTM}
	\acro{UniLSTM}[UniLSTM]{unidirectional LSTM}
	\acro{WNG}[WNG]{white noise gain}
	\acro{RIR}[RIR]{room impulse response}
	\acro{RTF}[RTF]{room transfer function}
	\acro{ATF}[ATF]{acoustic transfer function}
	\acro{DPIR}[DPIR]{direct-path impulse response}
	\acro{MVDR}[MVDR]{minimum variance distortionless response}
	\acro{LCMV}[LCMV]{linear-constraint minimum-variance}
	\acro{PMWF}[PMWF]{parametric multichannel wiener filter}
	\acro{GSC}[GSC]{Generalized sidelobe canceller}
	\acro{JNF}[JNF]{joint spatial and temporal-spectral nonlinear filtering}
	\acro{FT-JNF}[FT-JNF]{joint spatial and temporal-spectral nonlinear filtering}
	\acro{DSB}[DSB]{delay-and-sum beamformer}
	\acro{SDR}[SDR]{signal-to-distortion ratio}
	\acro{noisySDR}[reference microphone]{\ac{SDR} of the unprocessed omnidirectional reference microphone}
	\acrodefplural{noisySDR}[$\textrm{SDRs}^\textrm{omni}$]{\acp{SDR} of the unprocessed omnidirectional microphone}
	\acro{SNR}[SNR]{signal-to-noise ratio}
	\acro{STFT}[STFT]{short-time Fourier transform}
	\acro{MAE}[MAE]{mean absolute error}
	\acro{TF}[TF]{time-frequency}
	\acro{tsdr}[BA-$\varepsilon$-tSDR]{batch-aggregated and regularized thresholded \ac{SDR}}
	\acro{STOI}[STOI]{short term objective intelligibility}
	\acro{PESQ}[PESQ]{perceptual evaluation of speech quality}
	\acro{UCA}[UCA]{uniform circular array}
	\acro{NDF}[NDF]{neural directional filtering}
	\acro{DF}[DF]{directivity factor}
	\acro{DI}[DI]{directivity index}
	\acro{HRTF}[HRTF]{head-related transfer function}
	\acro{ILD}[ILD]{interaural level difference}
	\acro{FiLM}[FiLM]{feature-wise linear modulation}
	\acro{VDM}[VDM]{virtual directional microphone}
    \acro{cIRM}[cIRM]{complex ideal ratio mask}
    \acro{DRR}[DRR]{direct-to-reverberation ratio} 
\end{acronym}

\section{\label{sec:1} Introduction}
% A fixed beamformer captures sound with a prescribed spatial response---its directivity pattern---which determines how strongly each source is weighted according to its direction of arrival \citep{van1988beamforming}. Combining multiple beamformers with complementary directivity patterns enables spatial sound capture, preserving the spatial cues of the sound field. For example, first-order Ambisonic signals can be obtained from the outputs of first-order \acp{DMA} \citep{4538682}, using a compact array of omnidirectional microphones without requiring dedicated devices such as a Soundfield microphone or an Eigenmike \citep{meyer2002eigenmike}. This illustrates that sound capture with controlled directivity patterns provides a flexible solution for spatial sound recording.

Capturing sound with a controlled directivity pattern is central to many spatial audio applications. A directivity pattern determines how strongly a source is weighted according to its direction of arrival~\citep{van1988beamforming}, and different applications demand different patterns, e.g., highly directive, frequency-invariant, or steerable. It is well established that useful patterns can be synthesized from microphone arrays: first-order Ambisonic signals, for instance, can be obtained from a compact planar microphone array~\citep{4538682}. Classically, such patterns are realized through fixed beamforming, whose attainable directivity and robustness are tightly bound to the array size and the number of microphones, a key limitation for the compact arrays preferred in practice.

Fixed beamforming employs data-independent linear filters to realize a prescribed, time-invariant directivity pattern~\citep{brandstein2001microphone, benesty2018fixed}. The performance of fixed beamformers is typically evaluated in terms of their directivity pattern, \ac{WNG}, and \ac{DF}, where a higher \ac{DF} generally comes at the expense of a lower \ac{WNG}. At one extreme, \acp{DSB} maximize \ac{WNG} but provide a low \ac{DF}. At the other extreme, superdirective beamformers attain a high \ac{DF} at the cost of a
low \ac{WNG}~\citep{brandstein2001microphone}. \Acp{DMA}~\citep {elko1997steerable, elko2000superdirectional, Elko2004, benesty2012study,borra2020efficient} and \ac{LS} beamformers~\citep{ls_beamforming} lie between these extremes, trading off \ac{WNG} against \ac{DF}. In particular, \acp{DMA} amplify white noise at low frequencies when attempting to achieve highly directive patterns~\citep{benesty2012study}. Moreover, the maximum order of a \ac{DMA} is limited to $Q-1$ for a linear array~\citep{chen2014design} and $\lfloor (Q-1)/2 \rfloor$ for a circular array~\citep{huang2020continuously}, where $Q$ is the number of microphones. This limitation substantially restricts the spatial filtering capability of \acp{DMA}. \ac{LS} beamformers can approximate a desired directivity pattern while ensuring a specified minimum \ac{WNG}~\citep{mabande2009robust}, but significant deviations from the desired pattern can occur when the number of microphones is small or the desired directivity is high. Overall, achieving a highly directive pattern with fixed beamformers often requires a large number of microphones and a sufficiently large array~\citep{benesty2018fixed}.

Unlike fixed beamforming, parametric spatial filtering \citep{tashev2005microphone, kallinger2009spatial, thiergart2013geometry, thiergart2013informed, thiergart2014informed, 7038281, chakrabarty2017bayesian, 8683515} offers a data-dependent approach to achieving a desired directivity pattern. Conventional parametric filters \citep{tashev2005microphone, kallinger2009spatial, thiergart2013geometry} employ a relatively simple signal model, where the direct sound is modeled as a single plane wave per time-frequency bin and the reverberant sound is modeled as a time-varying diffuse sound field \citep{jacobsen2000coherence}. These filters are typically computed using instantaneous estimates of model parameters, such as the \ac{DOA} and diffuseness of the sound. However, the single-wave assumption is easily violated in practical scenarios \citep{thiergart2012sound}, resulting in inaccurate spatial capture and audible artifacts. To overcome these limitations, informed spatial filters \citep{thiergart2013informed, thiergart2014informed}, which unify classical beamforming and parametric filters, extend the signal model to include multiple plane waves per time-frequency bin. Although violations of the signal model are less likely to occur, these methods rely heavily on accurate multiple-source \ac{DOA} and diffuse-sound power estimation, which can be challenging, particularly in reverberant or multi-source environments containing non-speech signals \citep{thiergart2012sound}. Nevertheless, these methods offer valuable functionality in applications such as acoustic zooming \citep{thiergart2014acoustical} and automatic spatial gain control \citep{braun2014automatic}.

\chg{With the rise of deep learning in acoustics \citep{bianco2019machine}, numerous \ac{DNN}-based spatial filters have been proposed \citep{7472778,9413594,halimeh2022complex,gu2021complex, tesch_insights, ftjnf_steerable, 10889345}. Some studies \citep{7472778, 9413594, halimeh2022complex} compute multichannel masks and employ filter-and-sum processing. Others \citep{gu2021complex, tesch_insights,  ftjnf_steerable, 10889345} estimate a single-channel mask and apply it to a reference or selected microphone. These methods typically define a target angular region, preserving sources within it while suppressing all others, which yields a rectangular directivity pattern with an abrupt transition at the region boundary. As a result, sensitivity to directional errors increases, leading to abrupt changes in loudness near the boundary. These methods do not offer explicit control over the directivity pattern and mainly focus on noise reduction or speaker extraction. To study the capability of neural spatial filters, such as the \ac{FT-JNF} \citep{tesch_insights}, to extract and represent spatial information, the \ac{DSB} output has been used as the \ac{DNN} training target \citep{briegleb2023localizing, briegleb2024analysis}. This approach implicitly guides the \ac{DNN} to learn the directivity pattern of a \ac{DSB}. However, the \ac{DSB} usually has a frequency-variant directivity pattern and limited directivity at low frequencies.}

Recently, we proposed \ac{NDF} to enable explicit control over the directivity pattern for spatial filtering \citep{ndf_iwaenc}. The \ac{NDF} uses a mask-based nonlinear filter to approximate the input–output behavior of an ideal directional microphone with a predefined directivity pattern by minimizing a signal-level loss between the network output and the directional-microphone target. This formulation is designed to preserve a desired spatial response across directions, decoupling the pattern-control goal from incidental signal-enhancement effects such as noise reduction or speaker extraction. The preliminary study \citep{ndf_iwaenc} demonstrated that \ac{NDF} can approximate a $3^{\textrm{rd}}$-order \ac{DMA} pattern with a four-microphone circular array in anechoic conditions, thereby exceeding the order limit of conventional \acp{DMA}. In subsequent work \citep{huang2025steerable}, steerability was achieved by one-hot encoding the steering angle, which restricts steering to a predefined set of discrete directions observed during training. Moreover, both studies were limited to anechoic conditions, lacked evaluation tools suited to nonlinear spatial filters, and left the underlying processing mechanisms of \ac{NDF} unexplained. 

This paper addresses these limitations and presents a comprehensive study of \ac{NDF}, extending it to reverberant environments and to continuously steerable directivity patterns. Beyond our preliminary studies \citep{ndf_iwaenc, huang2025steerable}, the main contributions are as follows.
1)~\emph{Continuous steerability}: We propose a \ac{FiLM}-based conditioning mechanism that enables arbitrary, continuous steering directions, including directions unseen during training. 2)~\emph{Training enhancements}: We propose a batch-aggregated normalized $L_{1}$ loss that outperforms the SDR-based loss used in our previous work \citep{ndf_iwaenc}, as well as a dedicated training and simulation strategy for reverberant environments. 3)~\emph{In-depth analysis}: We investigate the processing mechanisms of \ac{NDF}, including its ability to maintain frequency-invariant directivity patterns above the spatial aliasing frequency and its generalization to unseen non-speech and moving sources. 4)~\emph{Pattern controllability}: We demonstrate that \ac{NDF} flexibly realizes frequency-invariant directivity patterns with a compact four-microphone array, including $6^{\textrm{th}}$-order and arbitrary user-defined patterns. 5)~\emph{Evaluation methods}: We propose generalized methods to evaluate the directivity pattern and the \ac{DF} of any mask-based method, enabling separate analysis of the effects on the direct and reverberant components. Finally, we present an application of \ac{NDF} to stereo sound recording using a compact microphone array.

The remainder of this paper is organized as follows: Section~\ref{sec:problem} formulates the problem. Section~\ref{sec:pm} details the proposed method, and the corresponding evaluation methods are presented in Section~\ref{sec:performance_metrics}. Section~\ref{sec:exp_setup} outlines the experimental setup. Sections~\ref{sec:exp_anechoic} and \ref{sec:exp_rvb} present the experimental study conducted in anechoic and reverberant conditions, respectively. Section~\ref{sec:unseen} investigates the \ac{NDF} performance for previously unseen moving sources. \rev{Section~\ref{sec:limitations} summarizes the limitations of \ac{NDF}.} Finally, Section~\ref{sec:cls} concludes the paper.

\section{Problem formulation}\label{sec:problem}

\begin{figure}[t!]
\centering
\includegraphics[width=0.95\reprintcolumnwidth]{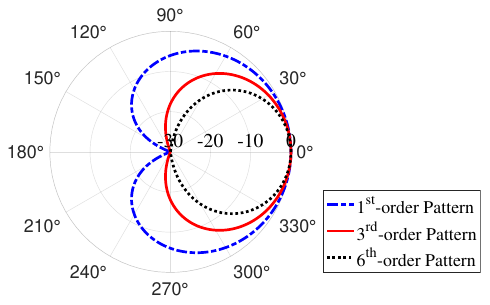} % double column width=0.95
\caption{\chg{Three directivity pattern examples on the $x$-$y$ plane. Steering direction $\theta_\textrm{s} = 0$ is used for the illustration.}}
\label{fig:dma_patterns}
\end{figure}

We consider a scenario in which a compact array with $Q$ omnidirectional microphones captures an acoustic scene comprising $N$ sound sources in the far field. Let $X_{q,n}[f,t]$ represent the $n$-th source signal at the $q$-th microphone in the \ac{STFT} domain, where $f$ and $t$ denote the frequency and time indices, respectively. The mixture signal at the $q$-th microphone, denoted by $Y_q[f,t]$, can be expressed as
\begin{equation} \label{eqn:mic_sig}
Y_q[f,t] = \sum_{n=1}^{N} X_{q,n}[f,t] + V_q[f,t],~q\in\{1,2,\ldots,Q\},
\end{equation}
where $V_q[f,t]$ represents the sensor noise that is spatially uncorrelated across the microphones. Furthermore, we have $X_{q,n}[f,t] = H_{\mathbf{p}_q,\mathbf{p}_n}[f] \, X_{n}[f,t]$ \citep{avargel2007multiplicative}, where $X_{n}[f,t]$ represents the $n$-th source signal and $H_{\mathbf{p}_q,\mathbf{p}_n}[f]$ models the \ac{ATF} between the $n$-th source at position $\mathbf{p}_n$ and the $q$-th microphone located at position $\mathbf{p}_q$.

The objective of the directional filtering task is to capture the acoustic scene and apply spatial filtering according to a specified directivity pattern. The directivity pattern describes the directional sensitivity of a beamformer or directional microphone, reflecting its spatial response to sounds arriving from various directions \citep{elko2000superdirectional, eargle2012microphone}. \chg{For example, a $1^\textrm{st}$-order \ac{DMA} directivity pattern \citep{elko2000superdirectional} is defined as
\begin{equation}
\Lambda_{1^\textrm{st}}(\theta, \phi) =\mu+ (1-\mu)(\sin \phi \sin   \phi_\textrm{s} \cos(\theta - \theta_\textrm{s}) + \cos \phi \cos   \phi_\textrm{s} ),
\end{equation}
where $\theta$ and $\phi$ represent the azimuth and polar angles of the incident sound, respectively. The parameter $\mu$, a real value in the interval $[0, 1]$, determines the null position; for instance, $\mu = 0.5$ yields a Cardioid pattern. Generally, a higher-order \ac{DMA} directivity pattern can be the product of multiple $1^\textrm{st}$-order \ac{DMA} patterns \citep{elko2000superdirectional}. In this study, we assume that all $1^\textrm{st}$-order \ac{DMA} patterns in the product are identical. This assumption ensures that the higher-order \ac{DMA} directivity pattern retains the same null positions as the $1^\textrm{st}$-order pattern and avoids additional sidelobes, since mainlobe control is the primary objective for sound capture. Therefore, a higher-order \ac{DMA} directivity pattern can be expressed as
\begin{equation}\label{eqn:simple_dma_pattern_definition}
\Lambda(\theta, \phi) = (\Lambda_{1^\textrm{st}}(\theta, \phi) )^{J},
\end{equation}
where $J$ is the order number. Figure~\ref{fig:dma_patterns} presents examples of $1^\textrm{st}$-, $3^\textrm{rd}$-, and $6^\textrm{th}$-order Cardioid directivity patterns, corresponding to $J \in \{1,3,6\}$. For these patterns, the mainlobe width decreases as the order increases.}

One possible approach for directional filtering is to mimic a \ac{VDM} with the desired directivity pattern. In the following, we assume that the \ac{VDM} position, denoted by $\mathbf{p}_\textrm{VDM}$, is equal to the position of the first microphone ($q=1$). The target signal for the directional filtering is the \ac{VDM} signal $Z[f,t]$ given by
\begin{equation}\label{eqn:vdm_sig_rvb}
Z[f,t] = \sum_{n=1}^{N} H_{\mathbf{p}_{\textrm{VDM}},\mathbf{p}_n}[f, \Lambda(\theta, \phi)] \, X_n[f,t],
\end{equation}
where $H_{\mathbf{p}_{\textrm{VDM}},\mathbf{p}_n}[f, \Lambda(\theta, \phi) ]$ denotes the \ac{RTF} between the $n$-th source at position $\mathbf{p}_n$ and the \ac{VDM}, which is given by
\begin{equation}\label{eqn:vdm_sig_h}
H_{\mathbf{p}_{\textrm{VDM}},\mathbf{p}_n}[f, \Lambda(\theta, \phi) ] = \sum_{i=1}^{\infty} \Lambda(\theta_i, \phi_i) \,  \rho^{(i)}_{\mathbf{p}_{\textrm{VDM}},\mathbf{p}_n}[f ] ,
\end{equation}
where $\rho^{(i)}_{\mathbf{p}_{\textrm{VDM}},\mathbf{p}_n}[f]$ represents the transfer function of the $i$-th sound propagation path between the $n$-th source and the \ac{VDM} in a reverberant environment. In other words, every reflection is weighted with the assigned gain based on the directivity pattern in the corresponding direction. Here, the incident angles $\theta_i$ and $\phi_i$ correspond to the angles of arrival of the $i$-th propagation path. \rev{In practice, Eq.~(\ref{eqn:vdm_sig_h}) is realized using the image source method\citep{RIRGenerator}, in which each propagation path $i$ corresponds to one image source. The position of the image source determines both the transfer function $\rho^{(i)}_{\mathbf{p}_{\textrm{VDM}},\mathbf{p}_n}[f]$ of that path and its incident angles $(\theta_i, \phi_i)$ at the \ac{VDM}. The directivity gain $\Lambda(\theta_i, \phi_i)$ is therefore applied to each path individually before the paths are summed, and the infinite sum is truncated at a finite reflection order.}
For simplicity, this paper focuses on a scenario where all sound sources are located in the $x$-$y$ plane, and we restrict the steering direction of the directivity pattern to the $x$-$y$ plane.

In an anechoic environment, there is only one direct-path transfer function $\rho_{\mathbf{p}_{\textrm{VDM}},\mathbf{p}_{n}}[f]$ between the $n$-th source and the \ac{VDM}, which simplifies Eq.~(\ref{eqn:vdm_sig_rvb}) as
\begin{equation}\label{eqn:vdm_sig_dir}
Z[f,t] = \sum_{n=1}^{N}  \Lambda(\theta_n)  \, \rho_{\mathbf{p}_{\textrm{VDM}},\mathbf{p}_n}[f] \, X_n[f,t],
\end{equation}
where $\theta_n$ represents the direction of arrival for the $n$-th source signal.  \ac{NDF} considers a \ac{DNN}-based approach to estimate a target \ac{VDM} signal using the microphone array signals.

\section{Proposed method}\label{sec:pm}
This section presents the proposed neural directional filtering method, which includes the \ac{DNN} architecture, loss function, and training strategy.

\begin{figure}[t!]
\centering
\includegraphics[width=0.95\reprintcolumnwidth]{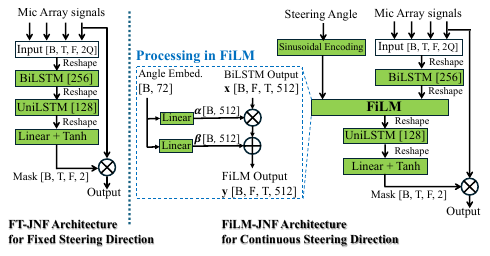} % double column width=0.95
\caption{\chg{DNN architecture for neural directional filtering: FT-JNF \citep{tesch_insights} for static steering direction; The proposed FiLM-JNF for continuous steering direction.}}
\label{fig:met_arch}
\end{figure}

\subsection{DNN architecture}
In this work, we adopt the \ac{FT-JNF} \citep{tesch_insights} as the \ac{DNN} architecture for the \ac{NDF} to learn a static directivity pattern, i.e., with a fixed steering direction (e.g., $\theta_s=0$). The architecture and intermediate feature map dimensions are shown on the left side of Figure~\ref{fig:met_arch}. In \ac{FT-JNF}, the real and imaginary parts of the $Q$ microphone signals in the \ac{STFT} domain are stacked along the channel dimension and then processed by two distinct \ac{LSTM} modules. The first \ac{LSTM} is a \ac{BiLSTM} operating on the stacked \ac{STFT} input along the frequency dimension. Its output is then processed by a \ac{UniLSTM} module. This module treats the frequency dimension as the batch dimension and processes along the temporal dimension, thereby modeling all frequencies independently and capturing the causal temporal relationships. Notably, unlike the two \acp{BiLSTM} in \citep{tesch_insights}, the unidirectional configuration of the second \ac{LSTM} enables causal processing. \rev{Subsequently, a linear layer with a hyperbolic tangent activation function computes a compressed complex-valued single-channel mask $\widetilde{\mathcal{M}}[f,t]$. Finally, the uncompressed mask $\mathcal{M}[f,t]$ is obtained using \citep{williamson2016complex} $\mathcal{M}[f,t] = \log\big[(1+\widetilde{\mathcal{M}}[f,t]) / (1-\widetilde{\mathcal{M}}[f,t])\big]$.} To use this \ac{DNN} to approximate the input-output behavior of a directional microphone in a signal-dependent manner, we compute an estimate for the target \ac{VDM} signal by masking the reference microphone signal:
\begin{equation}\label{eqn:dnn_op}
\widehat{Z}[f,t] = \mathcal{M}[f,t] Y_{1}[f,t].
\end{equation}

To enable continuous steering of the directivity pattern, we propose the FiLM-JNF architecture, which introduces a \ac{FiLM}~\citep{perez2018film} layer between the \ac{BiLSTM} and \ac{UniLSTM} layers of the \ac{FT-JNF} \citep{tesch_insights}, as illustrated on the right side of Figure~\ref{fig:met_arch}. In the proposed FiLM-JNF architecture, the desired steering direction is represented by an angle $\theta_{\mathrm{s}}$ (radians). We map $\theta_{\mathrm{s}}$ to an angle embedding $\mathbf{e}_{\theta_{\mathrm{s}}} \in \mathbb{R}^{d_{\mathrm{emb}}}$ with $d_{\mathrm{emb}}=72$, following the sinusoidal encoding \citep{vaswani2017attention} applied to the continuous angle instead of a discrete position. Concretely, for $i \in \{0,\ldots,d_{\mathrm{emb}}/2-1\}$, $\mathbf{e}_{\theta_{\mathrm{s}}}[2i] = \sin(\frac{\theta_{\mathrm{s}}}{10000^{2i/d_{\mathrm{emb}}}})$, $\mathbf{e}_{\theta_{\mathrm{s}}}[2i+1] = \cos(\frac{\theta_{\mathrm{s}}}{10000^{2i/d_{\mathrm{emb}}}})$, so that the batch of angles yields an embedding tensor of shape $[B, 72]$, which is then used to condition the network. The \ac{FiLM} layer computes per-feature affine parameters $\boldsymbol{\alpha}$ and $\boldsymbol{\beta}$ (dimensions $[B, 512]$) through two separate linear layers derived from the angle embeddings. It then applies element-wise modulation $\mathbf{y}=\boldsymbol{\alpha} \odot \mathbf{x} + \boldsymbol{\beta}$, shared across both time and frequency, where $\mathbf{x}$ denotes the \ac{BiLSTM} output features of dimension $[B, F, T, 512]$ and $\mathbf{y}$ is the output of the \ac{FiLM} layer. The output is reshaped to $[B \times F, T, 512]$ to fit the input requirements of the subsequent \ac{UniLSTM} layer. The remaining processing steps are identical to those in the \ac{FT-JNF}. 

\subsection{Loss function}\label{sec:loss}
\rev{A source-aggregated \ac{SDR} was proposed for source separation tasks and was shown to outperform the standard \ac{SDR} when used as a training loss~\citep{sasdr}. Since the target of the \ac{NDF} task is a single \ac{VDM} signal rather than a set of separated sources, an aggregation over sources is not applicable. Therefore, a modified version, the \ac{tsdr}, was adopted for training in~\citep{ndf_iwaenc}, in which $B$ samples in a mini-batch are aggregated; it is given by
\begin{equation}\label{eqn:tsdr}
\textrm{BA-$\varepsilon$-tSDR}(\mathbf{z},\widehat{\mathbf{z}}) =
10\log_{10}
\frac{\mathcal{E}}
{\sum_{b=1}^B \left\| \mathbf{z}^{(b)} - \widehat{\mathbf{z}}^{(b)} \right\|_{2}^{2} + \tau \mathcal{E}},
\end{equation}
where $\mathcal{E} = \sum_{b=1}^B \left\| \mathbf{z}^{(b)} \right\|_{2}^{2} + \varepsilon$ denotes the regularized energy of the target signals in the batch, $\mathbf{z}$ and $\widehat{\mathbf{z}}$ are the time-domain target and estimated \ac{VDM} signals, $\varepsilon$ is a small constant for numerical stability, and $\tau = 10^{-\frac{\textrm{SDR}_{\textrm{max}}}{10}}$ soft-thresholds the achievable SDR at
$\textrm{SDR}_{\textrm{max}} = 40$~\unit{\decibel}, which prevents the training from being dominated by samples that are already well estimated. Notably, the aggregation is performed over the $B$ samples of a mini-batch, which cover a diverse set of source directions, so that the \ac{tsdr} reflects the overall approximation quality over these directions rather than an average of per-sample \ac{SDR}; the latter would be dominated by samples whose target signal has very low energy, for example, when all sources in a sample mixture are located close to the null direction. Since a higher \ac{tsdr} indicates a better estimate, the loss function is defined as the negative \ac{tsdr}, which can be equivalently expressed as
\begin{equation}\label{eqn:tsdr-loss}
\begin{split}
\mathcal{L}_{\textrm{SDR}}(\mathbf{z},\widehat{\mathbf{z}})
&= -\,\textrm{BA-$\varepsilon$-tSDR}(\mathbf{z},\widehat{\mathbf{z}}) \\
&= 10\log_{10} \left(
\frac{\sum_{b=1}^B \left\| \mathbf{z}^{(b)} - \widehat{\mathbf{z}}^{(b)} \right\|_{2}^{2}}
{\sum_{b=1}^B \left\| \mathbf{z}^{(b)} \right\|_{2}^{2} + \varepsilon}
+ \tau \right).
\end{split}
\end{equation}}

It is often reported that the $L_{1}$ loss can outperform the $L_{2}$ loss for speech processing tasks in terms of metrics such as SDR, PESQ, and STOI \citep{7952120, pandey2018adversarial}. Therefore, we adopt a batch-aggregated normalized $L_{1}$ loss function in this work:
\begin{equation}\label{eqn:l1-loss}
\mathcal{L}_{\textrm{1}}(\mathbf{z},\widehat{\mathbf{z}}) =  \frac{\sum_{b=1}^B \left\| \mathbf{z}^{(b)} -  \widehat{\mathbf{z}}^{(b)} \right\|_{1}}{\sum_{b=1}^B \left\| \mathbf{z}^{(b)} \right\|_{1} + \varepsilon}.
\end{equation}
A performance comparison between the models trained with Eq.~(\ref{eqn:tsdr-loss}) and Eq.~(\ref{eqn:l1-loss}) is presented in Section~\ref{sssec:exp_anechoic_lossCompare}.

\subsection{Training strategy}\label{ssec:trainingStrategy}

\subsubsection{Training simulation for anechoic environment}\label{sssec:trainingStrategy-ane}
We set a fixed source-array distance $d$ for learning a far-field directivity pattern in the anechoic scenario, and assume the array to be placed at the origin of the coordinate system, and $P$ discrete candidate source positions are obtained by uniformly sampling the azimuth angle along a circle of radius $d$. The array and the source positions are assumed to be co-planar. We define a particular source-array setup as one acoustic scene. Within each scene, we randomly select $N$ positions from the $P$ source positions for $N$ speech sources. We then simulate direct-path transfer functions $\rho_{{\mathbf{p}_{\textrm{q}}},\mathbf{p}_{n}}[f]$ for all $Q$ microphones and $N$ sources using the \ac{RIR} generator \citep{RIRGenerator} with a reflection order of zero. Following this, we obtain $Q$ microphone signals using Eq.~(\ref{eqn:mic_sig}).

\subsubsection{Training simulation for reverberant environment}\label{sssec:trainingStrategy-rvb}
First, we randomly select $N$ \acp{DOA} for the $N$ sources from $P$ candidate source \acp{DOA}. To obtain a source-array setup, each source has a random source-array distance. Second, we define a room with a random size and a random reverberation time. Third, we randomly place the source-array setup in the room described in Sec.~\ref{sssec:setting_rvb}. The source-array setup lies in the room's $x$-$y$ plane. Finally, based on the current microphone and source positions, we generate the corresponding \acp{RIR} and compute the microphone signals. 

\subsubsection{Static or steerable}
For a static directivity pattern, we simulate one target \ac{VDM} signal $Z[f,t]$ with a fixed steering direction for each acoustic scene using Eq.~(\ref{eqn:vdm_sig_dir}) in anechoic conditions or using Eq.~(\ref{eqn:vdm_sig_rvb}) in reverberant conditions. For a steerable directivity pattern, we simulate $M$ target \ac{VDM} signals for steering directions uniformly spanning $0^{\circ}$ to $360^{\circ}$, where $M =\frac{360^{\circ}}{\vartheta}$ with $\vartheta$ denoting the angular resolution. The $m$-th \ac{VDM} target signal is also obtained using Eq.~(\ref{eqn:vdm_sig_rvb}) or Eq.~(\ref{eqn:vdm_sig_dir}) corresponding to the $m$-th steering direction. During training, we treat each microphone signal from an acoustic scene paired with a single \ac{VDM} target signal as a \emph{training sample}. When learning steerable directivity patterns, the same microphone signals are repeated $M$ times to train on $M$ \ac{VDM} target signals, yielding $M$ distinct training samples. Similarly, a \emph{test sample} is defined in the same way as the training sample.

\subsubsection{Mini-batch sampling}
\chg{Training samples with all sources near the null direction lead to excessively large losses, impacting stability. While the batch-aggregated loss helps, the problem persists if a mini-batch consists entirely of samples around the null direction.} Thus, we propose an enhanced mini-batch sampling strategy in which training samples are selected so that each mini-batch contains at least one example from the target direction or its vicinity ($\pm 20^\circ$). This prevents the loss in Eq.~(\ref{eqn:tsdr-loss}) or Eq.~(\ref{eqn:l1-loss}) from becoming excessively large, thereby improving training robustness.

\section{Performance measures}\label{sec:performance_metrics}
Since NDF targets the realization of a prescribed directivity pattern rather than noise reduction or source extraction, we evaluate it using metrics that characterize directional filtering behavior. For conventional linear beamformers, these are typically the directivity pattern and \ac{DF}. Since the \ac{NDF} is data-dependent and nonlinear, however, such metrics cannot be applied directly. We therefore propose an estimation procedure for the directivity pattern and the \ac{DF} that is applicable to nonlinear processing, where the two quantities characterize the spatial filtering of direct and reverberant sounds, respectively.

% The performance of conventional linear beamformers is commonly evaluated using the \ac{WNG}, \ac{DF}, and directivity pattern. As the \ac{NDF} is both data-dependent and nonlinear, we propose a method to estimate the directivity pattern and the \ac{DF} that is suitable for nonlinear processing methods. These analyze the spatial filtering of direct and reverberant sounds, respectively. 

To introduce the calculation of the proposed performance metrics, we let $X^{(k)}_{1, n}[f, t]$ be the \ac{STFT} representation of the $n$-th source signal in the $k$-th test sample at the reference microphone. In a reverberant environment, $X^{(k)}_{1, n}[f, t]$ can be decomposed as
\begin{equation}
X^{(k)}_{1, n}[f, t] = X^{(k)}_{1, n, \textrm{dir}}[f, t] + X^{(k)}_{1, n, \textrm{rvb}}[f, t],
\end{equation}
where $X^{(k)}_{1, n, \textrm{dir}}[f, t]$ represents the direct-path component and $X^{(k)}_{1, n, \textrm{rvb}}[f, t]$ represents the reverberant component (including all reflections) related to the $n$-th source. Consequently, we have $Y^{(k)}_{1, \textrm{dir}}[f, t] = \sum_{n=1}^{N} X^{(k)}_{1, n, \textrm{dir}}[f, t]$ and $Y^{(k)}_{1, \textrm{rvb}}[f, t] = \sum_{n=1}^{N} X^{(k)}_{1, n, \textrm{rvb}}[f, t]$, which represent the cumulative direct and reverberant components at the reference microphone, respectively.

\subsection{Directivity pattern}\label{ssec:performance_metrics_bp}
A directivity pattern describes the spatial responses of a spatial filter or directional microphone to sounds from different directions. In the following, we focus on estimating the power pattern, which equals the squared magnitude of the directivity pattern \citep{trees2002optimum}.

To estimate the power pattern obtained by a specific model, we apply the estimated mask $\mathcal{M}^{(k)}[f, t]$ for the $k$-th test sample separately to the direct-path part of each source signal as received by the reference microphone. The corresponding narrowband power ratio $\xi_{n}^{(k)}[f]$ of the masked source signals to the unmasked source signals is then calculated as
\begin{equation}
\xi_{n}^{(k)}[f] = \frac{ \sum_{t=1}^{T} \left| \mathcal{M}^{(k)}[f, t] \; X^{(k)}_{1, n, \textrm{dir}}[f, t] \right|^2}{\sum_{t=1}^{T}\left| X^{(k)}_{1, n, \textrm{dir}}[f, t] \right|^2},
\end{equation}
and the wideband power ratio $\bar{\xi}_{n}^{(k)}$ is given as
\begin{equation}
\bar{\xi}_{n}^{(k)} = \frac{ \sum_{f=1}^{F} \sum_{t=1}^{T} \left| \mathcal{M}^{(k)}[f, t] \; X^{(k)}_{1, n, \textrm{dir}}[f, t] \right|^2}{\sum_{f=1}^{F} \sum_{t=1}^{T}\left| X^{(k)}_{1, n, \textrm{dir}}[f, t] \right|^2},
\end{equation}
where $T$ represents the number of time frames and $F$ denotes the number of frequency bins. It should be noted that the mask is computed from the reverberant input and applied only to the direct sound. Therefore, the power ratio is more accurate when the \ac{DRR} is high.

\rev{After obtaining the power ratios, the power pattern for the \ac{NDF} model is estimated using the entire test set: each source is associated with a direction. For each direction, we average the power ratios in the logarithmic domain, which is equivalent to the geometric mean and less sensitive to outliers \cite{6054_mean}. %This is the same averaging used for segmental SNR in speech-enhancement evaluation \cite{hansen1998effective}. 
Mathematically, the narrowband power pattern for angle $\theta_p$ and frequency $f$ is given by
\begin{equation}
\widehat{\mathcal{P}}[\theta_p, f] =\frac{10}{|\mathcal{H}_{\theta_p}|}  \sum_{{(k, n)} \in \mathcal{H}_{\theta_p}} \log_{10}\xi_{n}^{(k)}[f]  ,
\end{equation}
where $\theta_p$ with $p \in \{1,2,\ldots,P\}$ is one of $P$ candidate source \acp{DOA} contained in the test dataset. Similarly, the wideband power pattern $\widehat{\mathcal{P}}[\theta_p]$ is given by
\begin{equation}
\widehat{\mathcal{P}}[\theta_p] =\frac{10}{|\mathcal{H}_{\theta_p}|} \sum_{(k,n) \in \mathcal{H}_{\theta_p}} \log_{10} \bar{\xi}_{n}^{(k)},
\end{equation}}
where $\mathcal{H}_{\theta_p}$ is a set of indices ($k$, $n$) that include all sources in the test dataset that are located in the direction $\theta_p$, i.e.,
\begin{equation}
\mathcal{H}_{\theta_p} = \left\{ (k, n) \mid \theta^{(k)}_{n} = \theta_p \right\},
\end{equation}
and $|\mathcal{H}_{\theta_p}|$ represents the cardinality of the set $\mathcal{H}_{\theta_p}$.

\subsection{Directivity factor}
The original definition of \ac{DF} describes a fixed beamformer's ability to suppress a diffuse noise field, and it is defined \citep{brandstein2001microphone} as
\begin{equation}
\widehat{\mathcal{DF}}_{\textrm{original}} =  \frac{ \left|\mathbf{w}^{H}\mathbf{d} \right|^2}{ \mathbf{w}^{H} \boldsymbol{\Gamma}\mathbf{w}   },
\label{eq:Df-old}
\end{equation}
where $\mathbf{w}$ denotes the weights of the conventional beamformer under test, $\mathbf{d}$ is the steering vector of the beamformer, and $\boldsymbol{\Gamma}$ is the spatial coherence matrix for a diffuse noise field. It is often assumed that the late reverberation can be modeled as a diffuse sound field. Consequently, the \ac{DF} is a measure for the amount of reverberation reduction.

If the beamformer is assumed to be distortionless~\citep{trees2002optimum}, i.e., $\left|\mathbf{w}^{H}\mathbf{d} \right|^2 = 1$ , then Eq.~(\ref{eq:Df-old}) can be written as
\begin{equation}
\widehat{\mathcal{DF}}_{\textrm{original}} =  \frac{1}{ \mathbf{w}^{H} \boldsymbol{\Gamma}\mathbf{w}}
=  \frac{\psi}{ \mathbf{w}^{H} \psi \, \boldsymbol{\Gamma}\mathbf{w}},
\label{eq:Df-old1}
\end{equation}
where $\psi$ is the diffuse noise power at the (unprocessed) first microphone, and $\mathbf{w}^{H} \psi \, \boldsymbol{\Gamma}\mathbf{w}$ is the diffuse noise power at the output.

In the following, we deliberately define the DF to characterize only the processing of the reverberant component. The response to the direct sound---including any attenuation or distortion along the steering direction---is assessed separately by the directivity pattern in Sec.~\ref{ssec:performance_metrics_bp}. Consequently, we should always consider both the directivity pattern and the DF to assess the effective reduction in reverberation. Accordingly, we propose the computation method for \ac{DF} as below
\begin{equation}
\widehat{\mathcal{DF}}\left[f \right]  =  \frac{ \sum_{k=1}^{K}   \sum_{t=1}^{T}\left| Y^{(k)}_{1, \textrm{rvb}}[f, t] \right|^2}{\sum_{k=1}^{K}    \sum_{t=1}^{T}\left| \mathcal{M}^{(k)}[f, t] Y^{(k)}_{1, \textrm{rvb}}[f, t] \; \right|^2},
\label{eq:Df-2}
\end{equation}
where $K$ is the number of test samples. The right-hand side of Eq.~(\ref{eq:Df-2}) describes the ratio of the power of the reverberant components at the input to that at the output, reflecting the mask's suppression of reverberation. It is worth noting that the directivity factor is estimated only from the reverberant component, and the mask is computed from the entire microphone signals. Therefore, the \ac{DF} is more accurate when the reverberant component dominates the microphone signal, i.e., when the \ac{DRR} is low.

In addition, we can obtain an estimation of \ac{DF} for the target \ac{VDM} signal using
\begin{equation}
\widehat{\mathcal{DF}}_{\mathrm{target}}\left[f \right]  =   \frac{  \sum_{k=1}^{K} \sum_{t=1}^{T} \left|Y^{(k)}_{1, \textrm{rvb}}[f, t]\right|^2 }{  \sum_{k=1}^{K}  \sum_{t=1}^{T}\left| Z^{(k)}_{\textrm{rvb}}[f,t] \right|^2} ,
\label{eq:Df-4}
\end{equation}
where $Z^{(k)}_{\textrm{rvb}}[f,t] $ is the reverberant component of the target \ac{VDM} signal for the $k$-th test sample.

\chg{\subsection{Signal estimation quality}
We use the \ac{SDR} \citep{vincent2006performance}, SCOREQ \citep{ragano2024SCOREQ}, and \ac{PESQ} \citep{pesqc2}, averaged over the test set, to measure the estimated signals' quality compared to the target \ac{VDM} signals.}

\section{Experimental setup}\label{sec:exp_setup}
This section provides a detailed description of the experimental setup, encompassing the array geometry, the target \ac{DMA} directivity patterns, the datasets, and the training details.

\subsection{Array geometry and DMA directivity patterns}\label{ssec:arrayanddmay}
We employed a four-microphone array ($Q=4$) comprising three microphones arranged in a \ac{UCA} and an additional microphone at the array center. Open circular arrays are known to suffer from deep-null problems at frequencies corresponding to zeros of the Bessel functions \citep{zhao2024circular}; the inclusion of a center microphone in the present configuration avoids this issue. In this paper, the center microphone serves as the reference microphone and coincides with the \ac{VDM} position. Hence, the target \ac{VDM} signal and the reference signal are inherently phase-aligned for any source direction, and the estimated mask only needs to implement the direction-dependent pattern gain without phase corrections. Unless stated otherwise, all models were trained and tested using a \ac{UCA} with a diameter of 3~\unit{\cm}. Furthermore, $1^\textrm{st}$-, $3^\textrm{rd}$-, and $6^\textrm{th}$-order Cardioid directivity patterns in Figure~\ref{fig:dma_patterns} were used to investigate \ac{NDF}.

\chg{
\subsection{Baselines}\label{sec:baselines}
\begin{figure}[t!]
\centering
\includegraphics[width=0.95\reprintcolumnwidth]{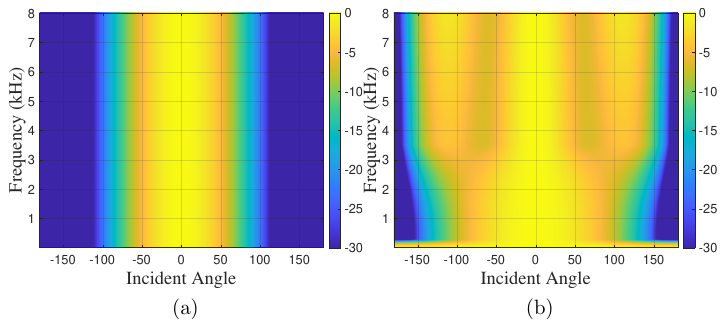}
\caption{(a) Optimization objective of the \ac{LS} beamformer. (b) Achieved pattern by the \ac{LS} beamformer with a minimum white noise gain constraint of $-15$~\unit{\decibel}.}
\label{fig:wls}
\end{figure}
To the best of our knowledge, fixed beamformers and parametric spatial filtering are the only effective spatial filtering methods for achieving the desired directivity pattern. Fixed beamformers \citep{elko2000superdirectional, brandstein2001microphone, benesty2018fixed} are designed to capture the sound field using a predefined directivity pattern. However, the achievable pattern is fundamentally limited by the array aperture and the number of microphones. For example, Figure~\ref{fig:wls} shows the resulting pattern obtained with a least-squares beamformer (LS beamformer) \citep{ls_beamforming} to target a $3^\textrm{rd}$-order Cardioid pattern, using the microphone array described in Section~\ref{ssec:arrayanddmay}. The LS beamformer incorporates a minimum \ac{WNG} constraint of $-15$~\unit{\decibel}. As shown, the LS beamformer does not achieve the desired frequency-invariant response, as it suffers from spatial aliasing at high frequencies and exhibits a wider mainlobe at low frequencies, thereby reducing spatial selectivity. As discussed in Section~\ref{sec:1}, with the array configuration described in Section~\ref{sec:exp_setup}, only the $1^\textrm{st}$-order pattern is achievable with conventional \ac{DMA}. Therefore, the LS beamformer \citep{ls_beamforming}  and a null-constraint \ac{DMA} \citep{huang2020continuously} are used as baselines for the $1^\textrm{st}$-order Cardioid pattern.

Alternatively, parametric directional filtering \citep{tashev2005microphone, kallinger2009spatial, thiergart2013geometry,thiergart2013informed,thiergart2014informed, 7038281, chakrabarty2017bayesian,8683515} can indeed approximate arbitrary directivity patterns. However, the performance of these approaches highly relies on the accuracy of the \ac{DOA} and coherence-to-diffuse power ratio estimates, which cannot be precisely obtained above the spatial aliasing frequency. To mitigate the effects of estimation errors, we use a simplified oracle parametric filter as our baseline for experiments in a simulated anechoic environment. Specifically, the parametric filter is computed using oracle DOA estimates that avoid potential artifacts from spatial aliasing, thereby providing an upper bound on its performance.
}

\subsection{Datasets}\label{ssec:datasets_details}
The training, validation, and test datasets were generated by convolving single-channel source signals with simulated \acp{RIR}.
The source signals for the training and validation sets were speech signals taken from the `train-clean-360' and `dev-clean' subsets of the LibriSpeech database \citep{librispeech}, respectively. Finally, all source signals were trimmed or zero-padded to a length of four seconds prior to convolution by the \ac{RIR}.

We used both speech and non-speech test sets to investigate the performance of \ac{NDF}. For the speech test sets, speech utterances were selected from the EARS dataset \citep{richter2024ears} with the criterion that their loudness is at least $-42$~LUFS \citep{loudness}. To achieve a relatively low proportion of silence within a speech segment, each utterance was then trimmed to a four-second segment that had a higher loudness level than the average loudness level of the original utterance. The non-speech test set used noise signals from the WHAM! dataset \citep{Wichern2019WHAM} as the source signals. If any sources were shorter than four seconds, we zero-padded them. However, for acoustic scenes containing multiple non-speech sources, the individual signals were trimmed to the length of the shortest source.

Similarly to \citep{ndf_iwaenc}, we normalized all convolved signals to have a loudness within $\left[-33, -25\right]$~LUFS. Additionally, we added white Gaussian noise to the array's microphone signals as self-noise. Unless otherwise specified, the \ac{SNR} for training and testing was $30$~\unit{\decibel} with respect to the mixture of all sources.

\subsubsection{Anechoic environment}\label{sssec:setting_anechoic}
We set a fixed source-array distance with $d = 1.5$~\unit{\m} for the anechoic environment.

\paragraph{Training datasets} We followed the training strategy described in Section~\ref{ssec:trainingStrategy}, and used the following parameters. The number of candidate source \ac{DOA}s for the training and validation sets was restricted to $P_\textrm{{train}} = 72$ with $\theta \in \{0^{\circ}, 5^{\circ}, \ldots, 355^{\circ}\}$ and $P_\textrm{{val}} = 72$ with $\theta \in \{2.5^{\circ}, 7.5^{\circ}, \ldots, 357.5^{\circ}\}$. For training a static directivity pattern with $\theta_s = 0$, the training and validation sets for a static directivity pattern consisted of $11520$ and $2880$ training samples, respectively. For training a steerable directivity pattern with potential steering directions with $\theta_\textrm{s}\in \{0^{\circ}, 5^{\circ}, \ldots, 355^{\circ}\}$, we generated $M = 72$ target \ac{VDM} signals for each scene, corresponding to a total of $1440 \times 72$ training samples in the training set and $360 \times 72$ training samples in the validation set.

\paragraph{Test datasets}
The number of candidate source \acp{DOA} for a test set was restricted to $P_\textrm{{test}} = 144$ with $\theta \in \{1.25^{\circ}, 3.75^{\circ}, \ldots, 358.75^{\circ}\}$. To ensure equal testing for each candidate source \ac{DOA}, we generated the test samples by uniformly sampling all candidate \acp{DOA}. Each test sample contained two concurrent sources. To test the models trained for a static directivity pattern, we generated $3240$ testing samples. \chg{To test the models trained for a steerable directivity pattern, we generated six target \ac{VDM} signals with $\theta_\textrm{s}\in \{0^{\circ}, 30^{\circ}, 32.5^{\circ},  60^{\circ}, 67.5^{\circ}, 90^{\circ}\}$.}

\subsubsection{Reverberant environment}\label{sssec:setting_rvb}
\begin{table}[t]
\setlength\extrarowheight{0.1pt}
\caption{Ranges for reverberant room acoustic settings.}
\begin{adjustbox}{max width=\columnwidth}
\begin{tabular}{ccccc}
\toprule
Length \ \ & Width \ \ & Height \ \ & $\textrm{RT}_{60}$ \ \ & Source-array dist. \\
\midrule
6 - 10~\unit{\metre} \ & 4 - 8~\unit{\metre} \ & 3 - 5~\unit{\metre} \ & 0.2 - 0.5~\unit{\s} \ & 0.5 - 2.5~\unit{\metre}  \\
\bottomrule
\end{tabular}
\end{adjustbox}
\label{tab:room_setting}
\end{table}
We simulated each reverberant training sample using the strategy described in Section~\ref{sssec:trainingStrategy-rvb} for training, validation, and test sets. The candidate source \ac{DOA}s for the training, validation, and test sets were the same as those in an anechoic environment setting. The source-array distance, room size (length, width, and height), and the $\textrm{RT}_{60}$ were uniformly sampled from the ranges in Table~\ref{tab:room_setting}. The array position in the room was chosen based on the Monte Carlo Room Impulse Response simulation \citep{MonteCarloRIR}, while ensuring that the sampled position is at least 1.2~\unit{\m} away from all walls. For the experimental study under reverberant conditions, we only trained the models with a static pattern. For a static pattern with $\theta_s = 0$, the training and validation sets consisted of 50000 and 6000 training samples, respectively. The test sets contained $3240$ test samples. Each test sample contained two concurrent sources.

\subsection{\chg{Training settings and complexity analysis}} \label{ssec:training_setting}
\begin{table}[t]
  \centering
  \caption{\chg{Model complexity and real-time factor (RTF): FT-JNF and FiLM-JNF. Python implementation with an ONNX model on an Apple MacBook Pro 2022 M2.}}
  \begin{adjustbox}{max width=\columnwidth}
    \begin{tabular}{lcccc}
      \toprule
      \textbf{Model} & \ \ \textbf{\#Parameters} & \ \ \textbf{Model Size} & \ \ \textbf{MACs/s} & \ \ \textbf{RTF} \\
        \midrule
      FT-JNF & 874~K & 3.33\,MB & 14.116\,G & 0.706 \\
      FiLM-JNF & 948~K & 3.62\,MB & 14.121\,G & 0.740 \\
      \bottomrule
    \end{tabular}
  \end{adjustbox}
  \label{tab:model-complexity}
\end{table}
Our earlier research, as described in \citep{ndf_iwaenc}, has shown that the \ac{NDF} model trained with two or more concurrently active sources can generalize to scenarios involving up to six sources. Since training with more than three sources did not significantly enhance the model's performance, we trained our models in this study using mixtures of up to three sources.

In anechoic environments, \ac{NDF} models for a static directivity pattern were trained for a maximum of $250$~epochs, while \ac{NDF} models for steerable directivity patterns or reverberant environments were trained up to $150$~epochs. The learning rate started at 0.001 and was multiplied by 0.75 every 40 epochs (anechoic) or 20 epochs (others). Training uses a batch size of $10$. The maximum null attenuation is limited to 30~\unit{\decibel}, with $\varepsilon$ in Eq.~(\ref{eqn:tsdr-loss}) and Eq.~(\ref{eqn:l1-loss}) set to $10^{-7}$.

In all the NDF models, the \ac{BiLSTM} layer contained $256$ hidden units, while the \ac{UniLSTM} layer contained $128$. \chg{The \ac{STFT} was computed on signal frames of $32$~\unit{\ms} duration, using a square-root Hann window with a $50\%$ overlap at a sampling frequency of $16$~\unit{\kilo\hertz}, resulting in $32$~\unit{\ms} algorithmic latency. For current settings, the complexity is analyzed in Table~\ref{tab:model-complexity}. The complexity for FiLM-JNF (14.121~G) was measured in multiply-accumulate operations (MACs) per second. \rev{All training parameters of the three training configurations are summarized in Table~\ref{tab:training_params}.}}
\begin{table}[t]
  \centering
  \caption{\rev{Training parameters of the \ac{NDF} models for the three training configurations. Values that apply to all configurations are given once.}}
  \label{tab:training_params}
  \renewcommand{\arraystretch}{1.15}
  \begin{adjustbox}{max width=\columnwidth}
    \begin{tabular}{l ccc}
      \toprule
      Parameter & \makecell{Anechoic\\(static)} & \makecell{Anechoic\\(steerable)} & \makecell{Reverberant\\(static)} \\
      \midrule
      Loss function & \multicolumn{3}{c}{batch-aggregated normalized $L_1$, Eq.~(\ref{eqn:l1-loss})} \\
      $\varepsilon$ in Eq.~(\ref{eqn:l1-loss}) & \multicolumn{3}{c}{$10^{-7}$} \\
      Maximum null attenuation & \multicolumn{3}{c}{30~\unit{\decibel}} \\
      \ac{SNR} (sensor noise) & \multicolumn{3}{c}{30~\unit{\decibel}} \\
      Mini-batch sampling & \multicolumn{3}{c}{$\geq 1$ sample within $\pm 20^{\circ}$ of the target direction} \\
      Concurrent sources per sample & \multicolumn{3}{c}{up to 3} \\
      Batch size & \multicolumn{3}{c}{10} \\
      Optimizer & \multicolumn{3}{c}{Adam} \\
      Initial learning rate & \multicolumn{3}{c}{0.001} \\
      Learning-rate decay & $0.75$/40~epochs & $0.75$/20~epochs & $ 0.75$/20~epochs \\
      Maximum epochs & 250 & 150 & 150 \\
      Training / validation samples & 11520 / 2880 & $1440 \times 72$ / $360 \times 72$ & 50000 / 6000 \\
      \bottomrule
    \end{tabular}
  \end{adjustbox}
\end{table}

\section{Evaluation in simulated anechoic environments}\label{sec:exp_anechoic}
In this section, we analyze the ability of \ac{NDF} models, trained in simulated anechoic environments, to learn the static \ac{DMA} patterns and explore mechanisms to achieve a frequency-invariant directivity pattern without spatial aliasing. Furthermore, we demonstrate the steerability of the models and their ability to learn user-defined patterns as well.

\subsection{Static DMA patterns}
Static pattern learning in an anechoic environment, by excluding additional challenges such as steerability and reverberation, provides an ideal experimental setup to explore the underlying processing mechanisms of \ac{NDF}.

\subsubsection{Loss function and baseline comparison}\label{sssec:exp_anechoic_lossCompare}

% \begin{table}[t]
%   \centering
%   \caption{\chg{SDR (\si{\decibel}, $\uparrow$ higher is better), Reference-based SCOREQ ($\downarrow$ lower is better), and PESQ ($\uparrow$ higher is better) for baseline methods and NDF with two loss functions.}}
%   \label{tab:loss_comparisons_sdr_SCOREQ_pesq}
%   \begin{adjustbox}{max width=\columnwidth}
%     \begin{tabular}{l ccc ccc ccc}
%       \hline\hline
%       & \multicolumn{3}{c}{$1^{st}$-order} & \multicolumn{3}{c}{$3^{rd}$-order} & \multicolumn{3}{c}{$6^{th}$-order} \\
%       \cline{2-4} \cline{5-7} \cline{8-10}
%       Method \ \ & SDR \ & SCOREQ \ & PESQ \ \ & SDR \ & SCOREQ  \ & PESQ \ \ & SDR \ & SCOREQ \ & PESQ \ \\
%       \hline
%       DMA \cite{huang2020continuously} & 6.25 & 1.10 & 2.33 & -- & -- & -- & -- & -- & -- \\
%       LS Beamformer \cite{ls_beamforming} & 10.32 & 1.22 & 2.14 & -- & -- & -- & -- & -- & -- \\
%       Parametric Filtering \cite{7038281} & 19.80 & 0.98 & 3.67 & 18.62 & 0.83 & 3.69 & 19.03 & 0.77 & 3.67 \\
%       \hline
%       NDF ($\mathcal{L}_{\mathrm{SDR}}$) \citep{ndf_iwaenc} & 27.55 & \textbf{0.89} & 4.43 & 25.71 & \textbf{0.74} & 4.39 & 25.68 & \textbf{0.69} & 4.38 \\
%       NDF ($\mathcal{L}_{1}$) & \textbf{27.70} & \textbf{0.89} & \textbf{4.45} & \textbf{26.93} & \textbf{0.74} & \textbf{4.42} & \textbf{27.31} & \textbf{0.69} & \textbf{4.41} \\
%       \hline\hline
%     \end{tabular}
%   \end{adjustbox}
% \end{table}

\begin{table}[t]
  \centering
  \caption{\chg{SDR (\si{\decibel}, $\uparrow$ higher is better), Reference-based SCOREQ ($\downarrow$ lower is better), and PESQ ($\uparrow$ higher is better) for baseline methods and NDF with two loss functions.}}
  \label{tab:loss_comparisons_sdr_SCOREQ_pesq}
  \renewcommand{\arraystretch}{1.15}
  \begin{adjustbox}{max width=\columnwidth}
    \begin{tabular}{l ccc @{\hskip 1.6em} ccc @{\hskip 1.6em} ccc}
      \toprule
      & \multicolumn{3}{c}{$1^{\textrm{st}}$-order} & \multicolumn{3}{c}{$3^{\textrm{rd}}$-order} & \multicolumn{3}{c}{$6^{\textrm{th}}$-order} \\
      \cmidrule(lr){2-4} \cmidrule(lr){5-7} \cmidrule(lr){8-10}
      Method & SDR & SCOREQ & PESQ & SDR & SCOREQ & PESQ & SDR & SCOREQ & PESQ \\
      \midrule
      DMA \cite{huang2020continuously} \ & 6.25 & 1.10 & 2.33 & -- & -- & -- & -- & -- & -- \\
      LS Beamformer \cite{ls_beamforming} \ & 10.32 & 1.22 & 2.14 & -- & -- & -- & -- & -- & -- \\
      Parametric Filtering \cite{7038281} \ & 19.80 & 0.98 & 3.67 & 18.62 & 0.83 & 3.69 & 19.03 & 0.77 & 3.67 \\
      \midrule
      NDF ($\mathcal{L}_{\mathrm{SDR}}$) \citep{ndf_iwaenc} \ & 27.55 & \textbf{0.89} & 4.43 & 25.71 & \textbf{0.74} & 4.39 & 25.68 & \textbf{0.69} & 4.38 \\
      NDF ($\mathcal{L}_{1}$) \ & \textbf{27.70} & \textbf{0.89} & \textbf{4.45} & \textbf{26.93} & \textbf{0.74} & \textbf{4.42} & \textbf{27.31} & \textbf{0.69} & \textbf{4.41} \\
      \bottomrule
    \end{tabular}
  \end{adjustbox}
\end{table}

\begin{figure}[t!]
\centering
\includegraphics[width=0.95\reprintcolumnwidth]{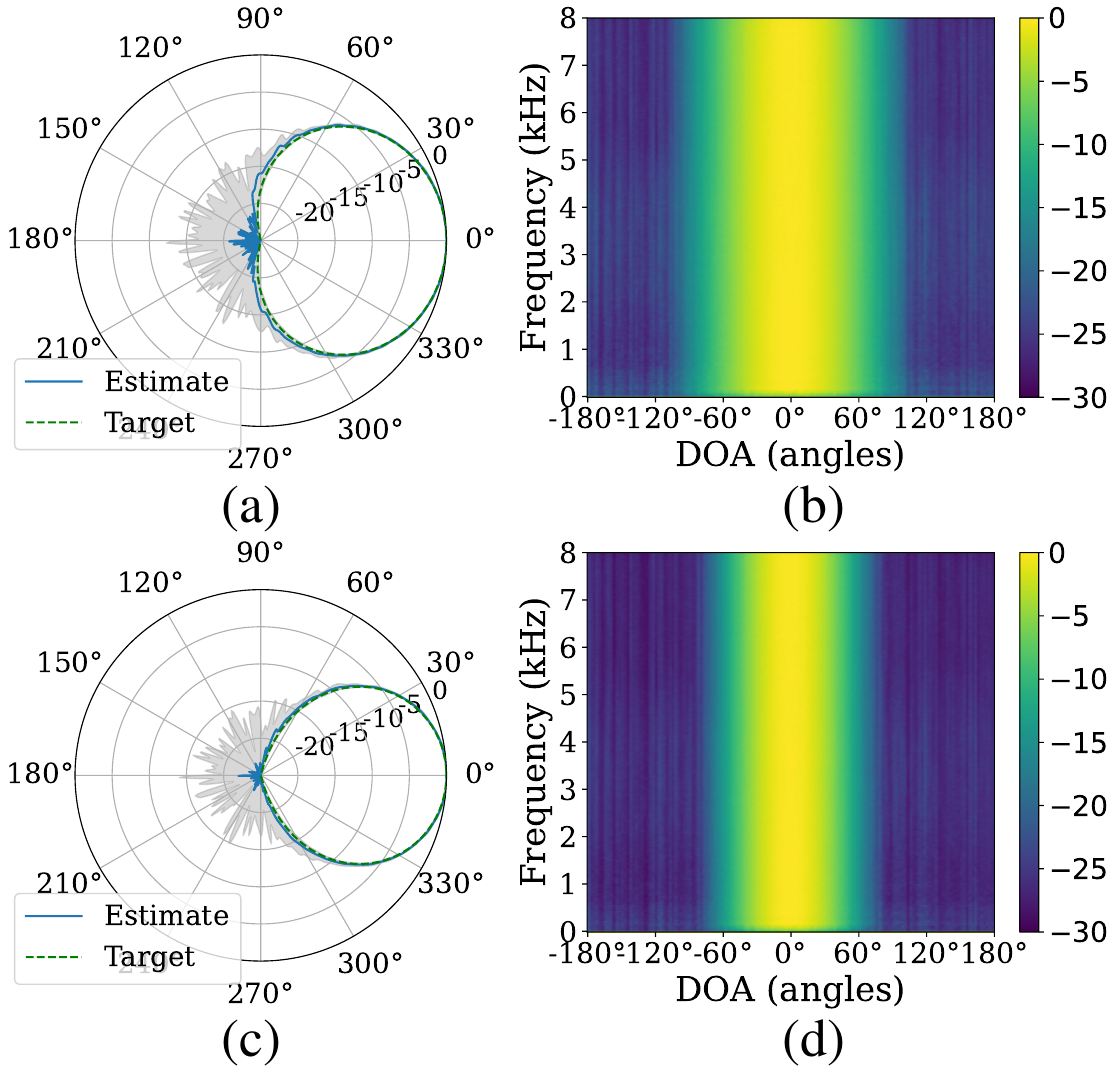} % double-cloumn: width=0.95
\caption{Estimated power patterns for the $3^{\textrm{rd}}$-order and $6^{\textrm{th}}$-order \ac{DMA} patterns. The diameter of the array is 3~cm. \rev{(a) $3^{\textrm{rd}}$-order, $\widehat{\mathcal{P}}[\theta]$; (b) $3^{\textrm{rd}}$-order, $\widehat{\mathcal{P}}[\theta,f]$; (c) $6^{\textrm{th}}$-order, $\widehat{\mathcal{P}}[\theta]$; (d) $6^{\textrm{th}}$-order, $\widehat{\mathcal{P}}[\theta,f]$. }The gray area in polar plots represents the standard deviation of the estimate.}
\label{fig:narrowband_bp}
\end{figure}

Table~\ref{tab:loss_comparisons_sdr_SCOREQ_pesq} shows the \chg{\ac{SDR}, SCOREQ, and \ac{PESQ}} performance of the \ac{NDF} models trained with the two loss functions described in Section~\ref{sec:loss} and the baseline systems (\ac{LS} beamformer and parametric filtering) as described in Section~\ref{sec:baselines}. We observe that the NDF models consistently outperform the baseline methods. The $3^{\textrm{rd}}$- and $6^{\textrm{th}}$-order patterns cannot be accurately approximated using the LS beamformer or DMA for the chosen compact array geometry as discussed in Section~\ref{sec:baselines}; hence, the corresponding entries are left blank, but the \ac{NDF} models can learn these higher-order patterns, as shown in Figure~\ref{fig:narrowband_bp}. Table~\ref{tab:loss_comparisons_sdr_SCOREQ_pesq} also shows that the \ac{NDF} model trained with the proposed batch-aggregated, normalized $\mathcal{L}_{\textrm{1}}$-loss function has better \ac{SDR} and \ac{PESQ} compared to the model trained with $\mathcal{L}_{\textrm{SDR}}$ for the $3^{\textrm{rd}}$- and $6^{\textrm{th}}$-order patterns, while the two models have similar performance for the $1^{\textrm{st}}$-order pattern.
Consequently, we use the $\mathcal{L}_{\textrm{1}}$ loss function for training the \ac{NDF} models in the following experiments.

\subsubsection{Power patterns and frequency processing mechanisms}\label{sec:freqProcMech}
\begin{figure}[t!]
\centering
\includegraphics[width=0.95\reprintcolumnwidth]{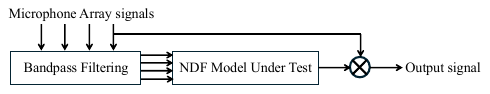}
\caption{Bandpass analysis of the \ac{NDF} models to study the frequency processing mechanisms.}
\label{fig:bandpass_diagram}
\end{figure}

\begin{figure}[t!]
\centering
\includegraphics[width=0.95\reprintcolumnwidth]{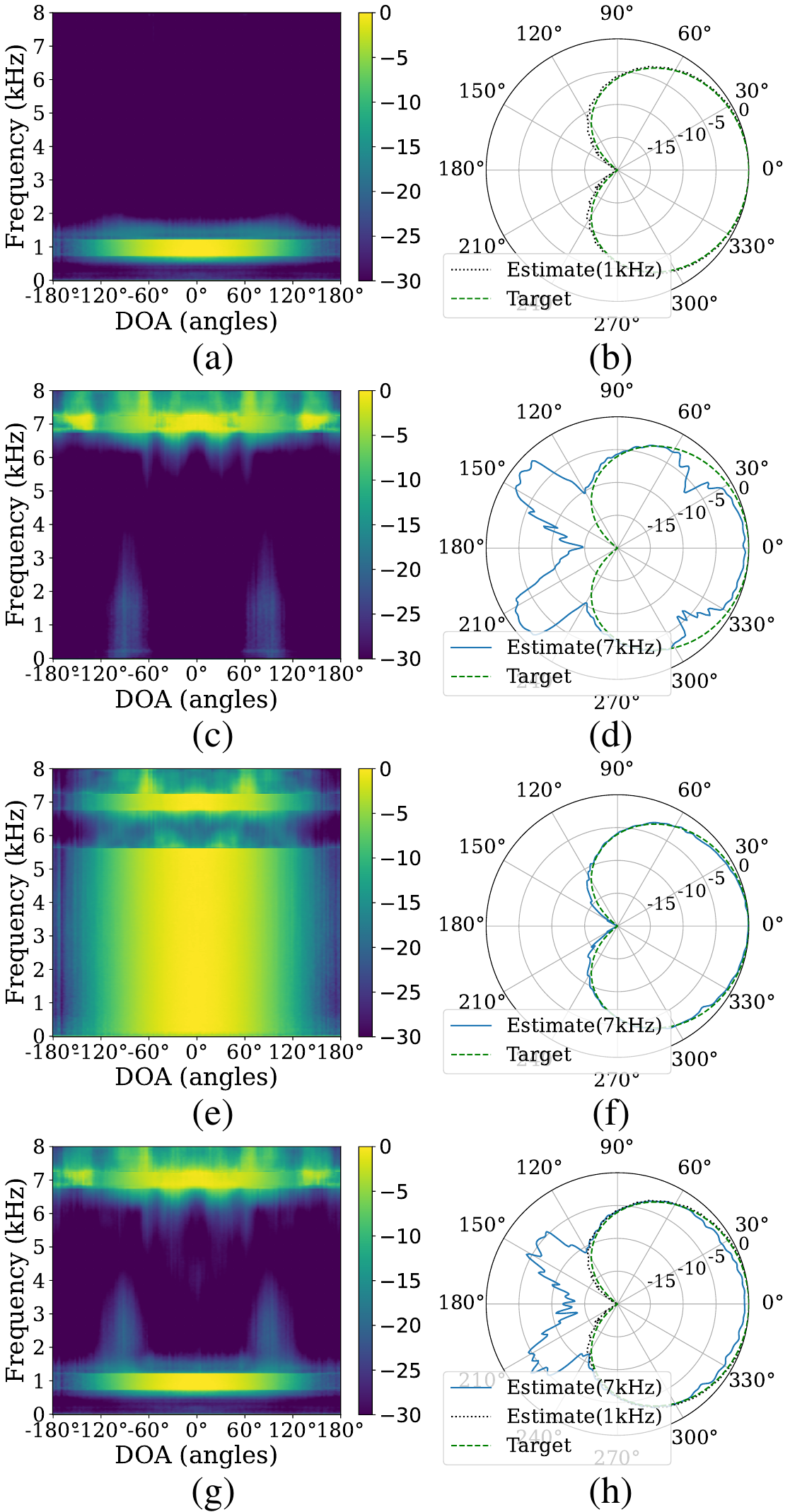} % double-cloumn: width=0.95
\caption{Bandpass analysis of an \ac{NDF} model trained for $1^{\textrm{st}}$-order pattern. The diameter of the array is 6~\unit{\cm}, where the spatial aliasing frequency corresponds to 5.6~\unit{\kilo\hertz}. \rev{(a) $\widehat{\mathcal{P}}[\theta, f]$ for a $1$~\unit{\kilo\hertz} bandpassed input; (b) $\widehat{\mathcal{P}}[\theta, f=1~\unit{\kilo\hertz}]$; (c) $\widehat{\mathcal{P}}[\theta, f]$ for a $7$~\unit{\kilo\hertz} bandpassed input; (d) $\widehat{\mathcal{P}}[\theta, f=7~\unit{\kilo\hertz}]$; (e) $\widehat{\mathcal{P}}[\theta, f]$ for a $7$~\unit{\kilo\hertz} bandpassed input together with the entire spectrum below $5.6$~\unit{\kilo\hertz}; (f) $\widehat{\mathcal{P}}[\theta, f=7~\unit{\kilo\hertz}]$; (g) $\widehat{\mathcal{P}}[\theta, f]$ for inputs bandpassed at $1$~\unit{\kilo\hertz} and $7$~\unit{\kilo\hertz}; (h) $\widehat{\mathcal{P}}[\theta, f=7~\unit{\kilo\hertz}]$ and $\widehat{\mathcal{P}}[\theta, f=1~\unit{\kilo\hertz}]$.}}
\label{fig:bp_plots}
\end{figure}

Figure~\ref{fig:narrowband_bp} shows the power pattern estimates for the $3^\textrm{rd}$- and $6^\textrm{th}$-order patterns. The \ac{NDF} effectively learns the mainlobe of these highly directive patterns, demonstrating strong spatial modeling capabilities. However, the null positions exhibit larger deviations. The narrowband results further indicate that the learned mainlobe patterns are largely frequency-invariant. This observation motivates a closer examination of the \ac{NDF} model's frequency processing mechanisms. Specifically, we aim to determine whether the model processes each frequency band primarily using local information or exploiting cross-band spectral structure. Furthermore, we investigate whether the observed frequency invariance of the learned patterns persists at frequencies well above the aliasing limit, thereby mitigating spatial aliasing effects. To this end, we employed a microphone array with a diameter of 6~\unit{\cm} for the subsequent experiments. With this configuration, spatial aliasing begins above 5.6~\unit{\kilo\hertz}, allowing us to test the model's performance under narrowband conditions both below and above this frequency. The model was trained and evaluated using data corresponding to this larger array.

We designed the experiment shown in Figure~\ref{fig:bandpass_diagram}. We used the speech test sets described in Section~\ref{ssec:datasets_details} and bandpass-filtered the corresponding microphone array signals before processing them with the \ac{NDF} model, trained on broadband speech signals. Based on preliminary studies, we set the bandwidth of the bandpass filter to 500~\unit{\hertz}. The mask provided by the \ac{NDF} model was then applied to the unprocessed reference microphone signal. This forced the \ac{NDF} model to use limited frequency bands.

As shown in Figures~\ref{fig:bp_plots}(a) and (b), when a bandpass signal centered at $1$~\unit{\kilo\hertz} is input into the \ac{NDF} model, the estimated patterns, using this narrowband spectral information, successfully match a desired $1^{\textrm{st}}$-order pattern. However, when a bandpass signal at $7$~\unit{\kilo\hertz} is provided, as shown in Figures~\ref{fig:bp_plots}(c) and (d), the \ac{NDF} model fails to approximate a target $1^{\textrm{st}}$-order pattern, rendering a deformed mainlobe due to spatial aliasing. In the following experiment, we provide the model with a signal featuring a band at 7~\unit{\kilo\hertz} and the entire spectral information below 5.6~\unit{\kilo\hertz}. Figures~\ref{fig:bp_plots}(e) and (f) show that the \ac{NDF} output no longer exhibits spatial aliasing at 7~\unit{\kilo\hertz} and effectively yields the target pattern. However, as shown in Figures~\ref{fig:bp_plots}(g) and (h), when the model receives a two-band signal at $1$~\unit{\kilo\hertz} and $7$~\unit{\kilo\hertz}, spatial aliasing appears at $7$~\unit{\kilo\hertz}.

These experiments show that the NDF model can effectively achieve a frequency-invariant pattern, and that information below the spatial aliasing frequency facilitates this. At higher frequencies where spatial aliasing occurs, the model appears to resolve ambiguities when broadband spectral context is available, suggesting cross-frequency processing. Leveraging low-frequency components to mitigate aliasing at high frequencies is found in classical signal processing; e.g., \citep{6868957} adopts a low-to-high subband multistage scheme, in which outputs from lower-frequency stages are propagated to higher-frequency stages to resolve high-frequency aliasing. The ability of \ac{FT-JNF} for spatial aliasing reduction is also found in the context of target speaker extraction in \citep{mannanova2025analysis}.

\begin{figure}[t!]
\centering
\includegraphics[width=0.95\reprintcolumnwidth]{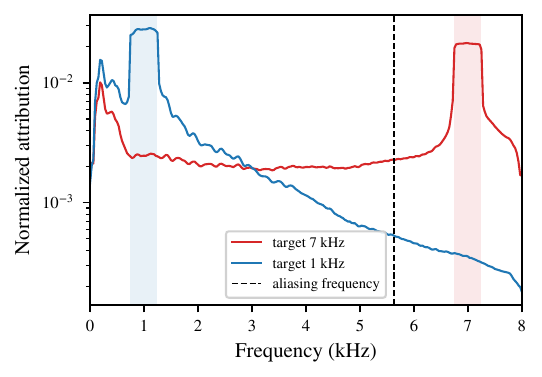}
\caption{\rev{Normalized attribution spectra of the $1^{\textrm{st}}$-order \ac{NDF} model, averaged over the complete single-source test set: contribution of each input frequency to the estimated mask at a target frequency above (7~\unit{\kilo\hertz}, red) and below (1~\unit{\kilo\hertz}, blue) the spatial aliasing frequency (dashed line). Shaded areas mark the $500$~\unit{\hertz} target bands $\mathcal{F}(f_{0}=1~\mathrm{kHz})$ and $\mathcal{F}(f_{0}=7~\mathrm{kHz})$. }}
\label{fig:attribution}
\end{figure}

%%% with new experiments
\rev{
So far, the above analysis only probes the model's input-output behavior. To quantify how much each input frequency contributes to the estimated mask at a given target frequency, we performed a gradient-based attribution analysis~\citep{simonyan2013deep} on single-source variants of the speech test set using the $1^{\textrm{st}}$-order model. Let $\breve{\mathbf{Y}}^{(k)} \in \mathbb{R}^{2Q \times F \times T}$ denote the stacked input tensor of the model for the $k$-th single-source test sample, obtained by stacking the real and imaginary parts of the $Q$-channel input \ac{STFT} $Y_q^{(k)}[f,t]$ along the channel dimension, and let $\widetilde{\mathcal{M}}^{(k)}[f,t]$ denote the compressed mask output of the network. We define a scalar score
\begin{equation}\label{eqn:attr-score}
s^{(k)}_{f_{0}} = \sum_{t=1}^{T} \sum_{f \in \mathcal{F}(f_{0})}
\big|\widetilde{\mathcal{M}}^{(k)}[f,t]\big|^{2},
\end{equation}
where $f_{0}$ denotes the target frequency and $\mathcal{F}(f_{0})$ represents the set of frequency bins within a $500$~\unit{\hertz} band centered at $f_{0}$. The score is defined on the compressed mask $\widetilde{\mathcal{M}}^{(k)}[f,t]$ rather than on $\mathcal{M}^{(k)}[f,t]$ for numerical stability: the decompression maps the bounded $\widetilde{\mathcal{M}}^{(k)}[f,t]$ to an unbounded $\mathcal{M}^{(k)}[f,t]$, and its derivative becomes arbitrarily large as $\widetilde{\mathcal{M}}^{(k)}[f,t]$ approaches $\pm 1$, which would corrupt the gradients. Subsequently, the normalized attribution spectrum is obtained by the gradient$\times$input method~\citep{shrikumar2016gradinput, ancona2017towards} as
\begin{equation}\label{eqn:attr-spectrum}
\begin{split}
\bar{A}^{(k)}_{f_{0}}[f] &= \frac{\sum_{t=1}^{T} A^{(k)}_{f_{0}}[f,t]}
{\sum_{f'=1}^{F} \sum_{t=1}^{T} A^{(k)}_{f_{0}}[f',t]}, \\
\textrm{with} \quad A^{(k)}_{f_{0}}[f,t] &= \sum_{c=1}^{2Q}
\left|
\frac{\partial s^{(k)}_{f_{0}}}{\partial \breve{Y}_c^{(k)}[f,t]}
\cdot \breve{Y}_c^{(k)}[f,t]
\right|.
\end{split}
\end{equation}
Here, $\bar{A}_{f_{0}}^{(k)}[f]$ represents the fraction of the total attribution to the estimated mask for the target frequency band $\mathcal{F}(f_{0})$ that originates from the input at frequency bin $f$; a large value indicates that the model relies strongly on this input frequency when estimating the mask at the target frequency. The attribution spectra are averaged over the complete single-source test set ($K = 3240$ samples covering all source directions), i.e., $\bar{A}_{f_{0}}[f] = \frac{1}{K}\sum_{k=1}^{K} \bar{A}_{f_{0}}^{(k)}[f]$, and the attribution mass of a frequency region is the sum of $\bar{A}_{f_{0}}[f]$ over the bins in that region. Figure~\ref{fig:attribution} shows the resulting attribution spectra for a target frequency above (7~\unit{\kilo\hertz}) and below (1~\unit{\kilo\hertz}) the aliasing frequency, and Table~\ref{tab:attribution} summarizes the corresponding distribution of the attribution mass. For the 7~\unit{\kilo\hertz} target, $44.7\,\%$ of the attribution mass lies below the aliasing frequency, forming a plateau that does not decay with spectral distance from the target. In contrast, the attribution of the 1~\unit{\kilo\hertz} target is concentrated around the target frequency and decays with spectral distance, with only $2.8\,\%$ contribution from frequencies above the aliasing frequency. Since both target frequencies are evaluated on the same inputs, the difference between the two attribution spectra cannot be explained by the spectral energy distribution of the input alone; this contrast shows that the model indeed draws on distant low frequencies when estimating the mask at 7~\unit{\kilo\hertz}, rather than merely on frequencies near the target frequency band. Repeating the analysis under the band-limited input condition of Figures~\ref{fig:bp_plots}(e) and (f) confirms this picture: when the input contains only the band around 7~\unit{\kilo\hertz} and the frequencies below the aliasing frequency, $57.7\,\%$ of the contribution to the 7~\unit{\kilo\hertz} mask estimate stems from the low frequencies, consistent with the recovery of the estimated patterns observed in this condition. Together with the bandpass experiments, these results provide converging behavioral and network-internal evidence that the model estimates the mask above the aliasing frequency by systematically drawing on the unaliased information below it.}

\begin{table}[t!]
  \centering
  \caption{\rev{Distribution of the attribution mass (gradient$\times$input) of the $1^{\textrm{st}}$-order \ac{NDF} model over three disjoint input-frequency regions, for a target frequency above (7~\unit{\kilo\hertz}) and below (1~\unit{\kilo\hertz}) the spatial aliasing frequency of 5.6~\unit{\kilo\hertz}. The target region comprises the set of frequency bins $\mathcal{F}(f_{0})$ centered at the target frequency $f_{0}$; the below- and above-aliasing regions exclude the target region.}}
  \label{tab:attribution}
  \renewcommand{\arraystretch}{1.25}
  \begin{adjustbox}{max width=\columnwidth}
    \begin{tabular}{l @{\hskip 1.6em} c @{\hskip 1.6em} c @{\hskip 1.6em} c}
      \toprule
      $f_{0}$ & Target region $\mathcal{F}(f_{0})$ & Below aliasing & Above aliasing \\
      \midrule
      7~\unit{\kilo\hertz} & 35.7\,\% & 44.7\,\% & 19.6\,\% \\
      1~\unit{\kilo\hertz} & 46.8\,\% & 50.4\,\% & \phantom{0}2.8\,\% \\
      \bottomrule
    \end{tabular}
  \end{adjustbox}
\end{table}

Although the underlying mechanism of these behaviors in the DNN model is not yet fully understood, \rev{two factors may underlie the cross-band information flow established above}. \rev{One hypothesis is that the model exploits spectral source characteristics:} cross-band dependencies in broadband sources, analogous to those in single-channel source separation, could provide contextual cues that help resolve ambiguous inter-microphone phase relationships at individual frequencies. \rev{A second hypothesis is that the model exploits the geometry of aliasing itself:} while inter-microphone phase differences no longer map uniquely to incident angles above the aliasing frequency, each angle still produces a distinct phase difference at each frequency, and the set of aliased angles varies with frequency while the true angle remains constant. Consequently, integrating phase information across frequencies \rev{could, in principle, geometrically} constrain the true angle of arrival. \rev{While the attribution analysis establishes that the model systematically draws on the low-frequency input, it does not by itself distinguish which of these two cues the model exploits.} We emphasize that these are hypothesized contributing factors; the observed robustness of the learned patterns above the aliasing frequency is an empirical finding rather than a theoretical guarantee that spatial aliasing has been resolved.

\subsubsection{Non-speech sources}
\begin{figure}[t!]
\centering
\includegraphics[width=0.95\reprintcolumnwidth]{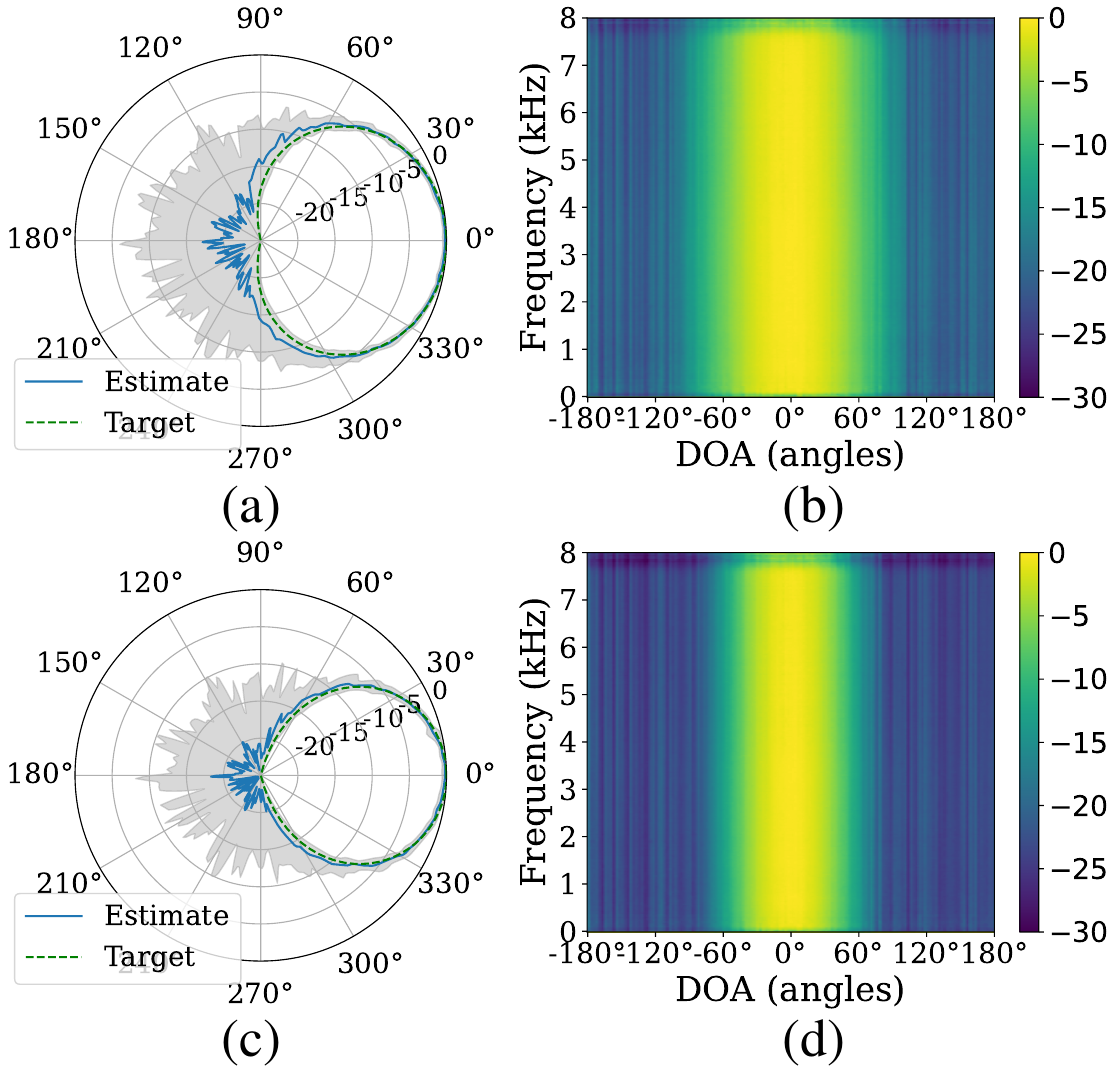} % double-cloumn: width=0.95
\caption{Evaluation of speech-trained \ac{NDF} models on the non-speech noise test set in anechoic conditions. \rev{Subfigures (a) and (b) illustrate the result for a $3^{\textrm{rd}}$-order target pattern, while subfigures (c) and (d) depict the results for a $6^{\textrm{th}}$-order target pattern.} The gray area in polar plots represents the standard deviation of the estimate.}
\label{fig:bp-wham}
\end{figure}

\begin{table}[t!]
  \centering
  \caption{\rev{SDR (\si{\decibel}) on the non-speech (WHAM!) test sets for the parametric filtering baseline and NDF. %PESQ and SCOREQ are omitted since they are designed for speech signals and are not applicable to non-speech sources.
  }}
  \label{tab:sdr_wham}
  \renewcommand{\arraystretch}{1.15}
  \begin{adjustbox}{max width=\columnwidth}
    \begin{tabular}{l cc}
      \toprule
      Method  \ \  \ & $3^{\textrm{rd}}$-order  \ \ & $6^{\textrm{th}}$-order \\
      \midrule
      Parametric Filtering \citep{7038281} \  \ \  & 12.82  \ \ & 14.22 \\
      NDF  \  \ \ & \textbf{18.34}  \ \ & \textbf{20.42} \\
      \bottomrule
    \end{tabular}
  \end{adjustbox}
\end{table}

%To verify whether \ac{NDF} models may generalize to signals with unseen spectral characteristics in the training, we evaluate the \ac{NDF} models on non-speech test sets defined in Section~\ref{ssec:datasets_details}. Figure~\ref{fig:bp-wham} shows the narrowband and wideband power patterns estimated using non-speech test sets for $3^{\textrm{rd}}$-order and $6^{\textrm{th}}$-order patterns. Although the deviations of the estimated patterns in Figure~\ref{fig:bp-wham} are greater than those observed in Figure~\ref{fig:narrowband_bp} (which was obtained using speech sources), the estimated power patterns maintain a good mainlobe approximation. This result demonstrates that speech-trained \ac{NDF} models can still perform directional filtering with the desired pattern, even for previously unseen non-speech noise sources. Thus, we conclude that the NDF models generalize to unseen sources during training. In other words, even when the spectral features of speech are absent, the NDF models can still extract and exploit the necessary spatial features based on the spectrum of non-speech signals.

To verify whether \ac{NDF} models may generalize to signals with unseen spectral characteristics in the training, we evaluated the \ac{NDF} models on non-speech test sets defined in Section~\ref{ssec:datasets_details}. Figure~\ref{fig:bp-wham} shows the narrowband and wideband power patterns estimated using non-speech test sets for $3^{\textrm{rd}}$-order and $6^{\textrm{th}}$-order patterns. Although the deviations of the estimated patterns in Figure~\ref{fig:bp-wham} are greater than those observed in Figure~\ref{fig:narrowband_bp} (which was obtained using speech sources), the estimated power patterns maintain a good mainlobe approximation. \rev{In addition to the power patterns, Table~\ref{tab:sdr_wham} reports the SDR obtained on the non-speech test sets; PESQ and SCOREQ are not reported, as both metrics are designed for speech signals. Compared with the speech test sets in Table~\ref{tab:loss_comparisons_sdr_SCOREQ_pesq}, the SDR of the NDF models decreases (e.g., from 26.93~\unit{\decibel} to 18.34~\unit{\decibel} for the $3^{\textrm{rd}}$-order pattern), which is consistent with the larger pattern deviations observed in Figure~\ref{fig:bp-wham}. Nevertheless, the NDF models still attain a relatively high SDR and outperform the parametric filtering baseline by more than 5~\unit{\decibel} for both the $3^{\textrm{rd}}$-order and $6^{\textrm{th}}$-order patterns. These results indicate that speech-trained \ac{NDF} models can still perform directional filtering with the desired pattern, even for previously unseen non-speech noise sources. This suggests that the models rely primarily on spatial cues rather than speech-specific spectral features. However, the performance degradation (larger pattern deviations and lower \ac{SDR}) shows that matching the spectral characteristics of the test and training signals still matters.

}

\subsubsection{Array aperture}
\begin{figure}[t!]
\centering
\includegraphics[width=0.95\reprintcolumnwidth]{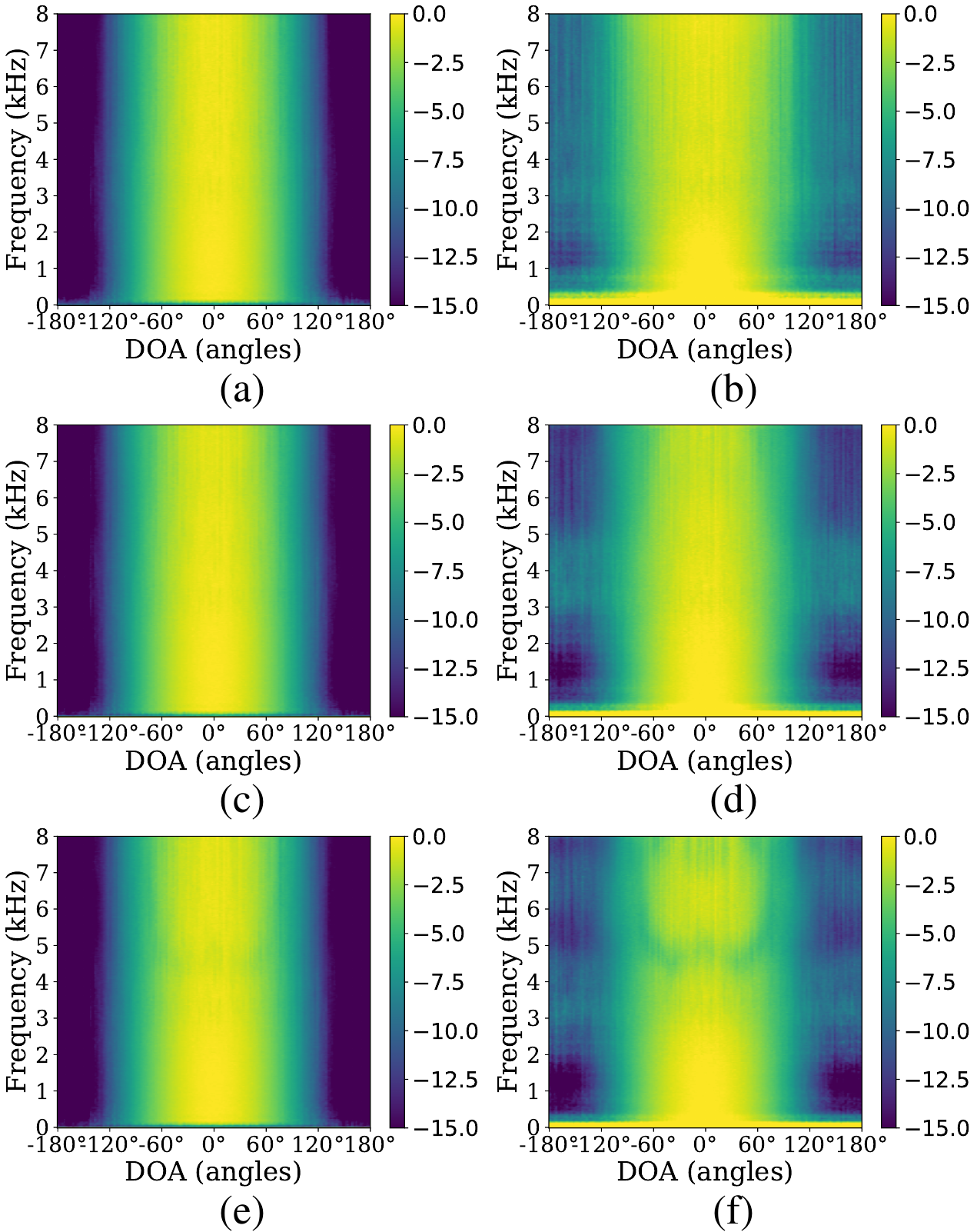} % double-cloumn: width=0.95
\caption{$\widehat{\mathcal{P}}[\theta, f]$ evaluation for different spatially uncorrelated sensor noise powers and array diameters $r$, for the \ac{NDF} models trained for the $1^{\textrm{st}}$-order pattern. \rev{(a) SNR = $20$~\unit{\decibel}, $r=3$~\unit{\cm}; (b) SNR = $10$~\unit{\decibel}, $r=3$~\unit{\cm}; (c) SNR = $20$~\unit{\decibel}, $r=6$~\unit{\cm}; (d) SNR = $10$~\unit{\decibel}, $r=6$~\unit{\cm}; (e) SNR = $20$~\unit{\decibel}, $r=9$~\unit{\cm}; (f) SNR = $10$~\unit{\decibel}, $r=9$~\unit{\cm}.}}
\label{fig:bp_diameters_snr}
\end{figure}

% \begin{table}[t]
%   \centering
%   \caption{\chg{SDR (\si{\decibel}) and PESQ of NDF models trained for arrays with different diameters (SNR = \SI{30}{\decibel}).}}
%   \label{tab:sdr_pesq_diameters_Cardioid}
%   \begin{adjustbox}{max width=\columnwidth}
%     \begin{tabular}{l cc cc cc}
%       \hline\hline
%       & \multicolumn{2}{c}{\SI{3}{\centi\meter}} & \multicolumn{2}{c}{\SI{6}{\centi\meter}} & \multicolumn{2}{c}{\SI{9}{\centi\meter}} \\
%       \cline{2-3} \cline{4-5} \cline{6-7}
%      Pattern \ \ \ & SDR \ & PESQ \ \ & SDR \ & PESQ \ \ & SDR & \ PESQ \\
%       \hline
%       $1^{st}$-order & 27.70 & 4.45 & 29.61 & 4.47 & \textbf{30.43} & \textbf{4.48} \\
%       $3^{rd}$-order & 26.93 & 4.42 & 28.85 & 4.45 & \textbf{29.45} & \textbf{4.46} \\
%       $6^{th}$-order & 27.31 & 4.41 & 28.93 & 4.44 & \textbf{29.49} & \textbf{4.45} \\
%       \hline\hline
%     \end{tabular}
%   \end{adjustbox}
% \end{table}

\begin{table}[t]
  \centering
  \caption{\rev{SDR (\si{\decibel}) and PESQ of \ac{NDF} models trained and tested for arrays with different diameters (SNR = 30~\unit{\decibel}).}}
  \label{tab:sdr_pesq_diameters_Cardioid}
  \renewcommand{\arraystretch}{1.15}
  \begin{adjustbox}{max width=\columnwidth}
    \begin{tabular}{l cc @{\hskip 1.6em} cc @{\hskip 1.6em} cc}
      \toprule
      & \multicolumn{2}{c}{\SI{3}{\centi\meter}} & \multicolumn{2}{c}{\SI{6}{\centi\meter}} & \multicolumn{2}{c}{\SI{9}{\centi\meter}} \\
      \cmidrule(lr){2-3} \cmidrule(lr){4-5} \cmidrule(lr){6-7}
      Pattern & SDR & PESQ & SDR & PESQ & SDR & PESQ \\
      \midrule
      $1^{\textrm{st}}$-order \ \ & 27.70 & 4.45 & 29.61 & 4.47 & \textbf{30.43} & \textbf{4.48} \\
      $3^{\textrm{rd}}$-order \ \ & 26.93 & 4.42 & 28.85 & 4.45 & \textbf{29.45} & \textbf{4.46} \\
      $6^{\textrm{th}}$-order \ \ & 27.31 & 4.41 & 28.93 & 4.44 & \textbf{29.49} & \textbf{4.45} \\
      \bottomrule
    \end{tabular}
  \end{adjustbox}
\end{table}
We investigated spatial aliasing in Section~\ref{sec:freqProcMech} for a \ac{UCA} with a diameter of 6~\unit{\cm}. Since the array diameter often affects the performance of fixed beamforming \citep{huang2020differential, yan2025neural}, we investigated how the array diameter affects the performance of the \ac{NDF} models. To this end, we trained \ac{NDF} models with array diameters of 3~\unit{\cm}, 6~\unit{\cm}, and 9~\unit{\cm}, and evaluated each model using test sets generated for the corresponding diameter. For both training and testing, the \ac{SNR} was set to 30~\unit{\decibel}.

Table~\ref{tab:sdr_pesq_diameters_Cardioid} shows that the \ac{SDR} and \ac{PESQ} improve as the diameter increases. This observation raises the question: Is a larger diameter always better? To investigate this further, we increased the microphone sensor noise in the test sets, reducing the \ac{SNR} to 20~\unit{\decibel} and 10~\unit{\decibel} while maintaining the models trained at an \ac{SNR} of 30~\unit{\decibel}. As depicted in Figure~\ref{fig:bp_diameters_snr}, using the $1^{\textrm{st}}$-order pattern as an example, we observe that at an SNR of $20$~\unit{\decibel}, the estimated patterns remain consistent across different diameters. At an \ac{SNR} of $10$~\unit{\decibel}, the \ac{NDF} renders an omnidirectional response at very low frequencies, and its response is actually larger than 0~\unit{\decibel} (e.g., up to 3.5~\unit{\decibel} for $r=3$~\unit{\cm}). This phenomenon is similar to the white-noise amplification issue observed in some fixed beamformers, such as \ac{DMA} and superdirective beamformers \citep{benesty2018fixed}. It is noted that the 3~\unit{\cm} diameter array has a more severe amplification problem than the 9~\unit{\cm} diameter array. However, as the diameter increases, particularly at $9$~\unit{\cm}, the \ac{NDF} model no longer maintains a frequency-invariant pattern at high frequencies for an \ac{SNR} of $10$~\unit{\decibel}. Therefore, under low \ac{SNR} conditions, a smaller diameter preserves the frequency-invariant shape of the estimated patterns at high frequencies. In comparison, a larger diameter enhances the robustness of low frequencies and exhibits better SDR.

\rev{\subsubsection{Source-array distances}   \label{sssec:sourceArrayDist} 
\begin{figure}[t!]
\centering
\includegraphics[width=0.95\reprintcolumnwidth]{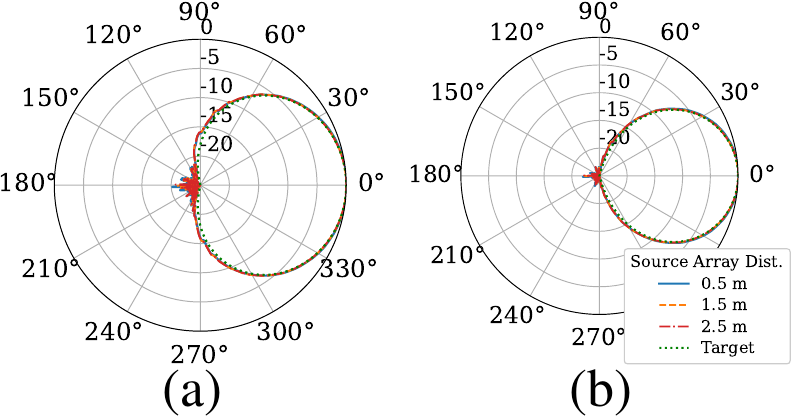} % double-cloumn: width=0.95
\caption{\rev{Estimated wideband power patterns of the NDF models across varying source-array distances. (a) $3^{\textrm{rd}}$-order pattern; (b) $6^{\textrm{th}}$-order pattern.}  }
\label{fig:sdr-distance}
\end{figure}

\begin{table}[t]
  \centering
  \caption{\rev{SDR (\si{\decibel}) for \ac{NDF} models trained at a source-array distance of 1.5~\unit{\m} and tested at varying source-array distances. Both training and testing use an array diameter of 3~\unit{\cm}.}}
  \label{tab:sdr_results_sourceArrayDist}
  \renewcommand{\arraystretch}{1.15}
  \begin{adjustbox}{max width=\columnwidth}
    \begin{tabular}{l ccccc}
      \toprule
      & \multicolumn{5}{c}{Source-array distance} \\
      \cmidrule(lr){2-6}
      Pattern & 0.5~\unit{\m} \ \ & 1.0~\unit{\m} \ \ & 1.5~\unit{\m}  \ \ & 2.0~\unit{\m}  \ \ & 2.5~\unit{\m} \\
      \midrule
      $3^{\textrm{rd}}$-order  \ \ \ & 22.87 & 26.61 & 26.93 & 26.84 & 26.68 \\
      $6^{\textrm{th}}$-order \ \ \ & 24.89 & 27.16 & 27.31 & 27.17 & 27.04 \\
      \bottomrule
    \end{tabular}
  \end{adjustbox}
\end{table}

In the experimental setup (Section~\ref{sssec:setting_anechoic}) and in our previous work \citep{ndf_iwaenc}, the default source-array distance for training and testing in an anechoic environment was set to $1.5$~\unit{\m}. Here, we investigate the generalizability of the \ac{NDF} models to varying source-array distances during inference. Table~\ref{tab:sdr_results_sourceArrayDist} reports the \ac{SDR} of models trained at a fixed distance of $1.5$~\unit{\m} and tested at distances ranging from $0.5$~\unit{\m} to $2.5$~\unit{\m}. The results indicate minor performance variation across the tested distances, except for lower \ac{SDR} observed at the notably close distance of $0.5$~\unit{\m}. We attribute this to the fact that, at such a close distance, the wavefront curvature deviates from the far-field propagation conditions under which the model was trained. Figure~\ref{fig:sdr-distance} shows the corresponding estimated wideband power patterns for the $3^{\textrm{rd}}$- and $6^{\textrm{th}}$-order models at distances of $0.5$~\unit{\m}, $1.5$~\unit{\m}, and $2.5$~\unit{\m}. The power patterns are estimated using the power ratios described in Section~\ref{ssec:performance_metrics_bp}, i.e., no normalization with respect to the steering direction is applied; hence, any overall gain deviation at unseen distances would be directly visible in the patterns. We observe that the estimated patterns are nearly identical across the tested distances: the response along the steering direction remains at approximately $0$~\unit{\decibel}, and the mainlobe shape is preserved.  Overall, these results show that as long as the far-field condition is satisfied, training a model at a fixed source-array distance is sufficient to capture sounds from varying source-array distances in anechoic environments.}

\subsection{Steerable DMA patterns}
\begin{table}[t!]
  \centering
  \caption{\rev{Performance of steerable \ac{NDF} models across various orders. }}
  \label{tab:sdr_ssf}
  \renewcommand{\arraystretch}{1.15}
  \begin{adjustbox}{max width=\columnwidth}
  \begin{tabular}{l ccc @{\hskip 1.6em} ccc @{\hskip 1.6em} ccc}
    \toprule
    & \multicolumn{3}{c}{$1^{\textrm{st}}$-order} & \multicolumn{3}{c}{$3^{\textrm{rd}}$-order} & \multicolumn{3}{c}{$6^{\textrm{th}}$-order} \\
    \cmidrule(lr){2-4} \cmidrule(lr){5-7} \cmidrule(lr){8-10}
    $\theta_\textrm{s}$ & SDR & SCOREQ & PESQ & SDR & SCOREQ & PESQ & SDR & SCOREQ & PESQ \\
    \midrule
    $0^{\circ}$   \ \ \  & 27.67 & 0.89 & 4.45 & 24.75 & 0.74 & 4.40 & 25.73 & 0.69 & 4.39 \\
    $30^{\circ}$   \ \  \  & 27.72 & 0.89 & 4.45 & 25.17 & 0.74 & 4.40 & 26.25 & 0.69 & 4.39 \\
    $32.5^{\circ}$  \ \ \ & 27.70 & 0.89 & 4.45 & 25.16 & 0.74 & 4.40 & 26.25 & 0.69 & 4.39 \\
    $60^{\circ}$   \ \ \    & 27.66 & 0.89 & 4.45 & 24.94 & 0.74 & 4.39 & 26.22 & 0.69 & 4.38 \\
    $67.5^{\circ}$  \ \ \ & 27.68 & 0.89 & 4.45 & 24.92 & 0.74 & 4.39 & 26.19 & 0.69 & 4.38 \\
    $90^{\circ}$  \ \ \    & 27.67 & 0.89 & 4.45 & 24.72 & 0.74 & 4.40 & 26.21 & 0.69 & 4.39 \\
    \bottomrule
  \end{tabular}
  \end{adjustbox}
\end{table}

\begin{figure}[t!]
\centering
\includegraphics[width=0.95\reprintcolumnwidth]{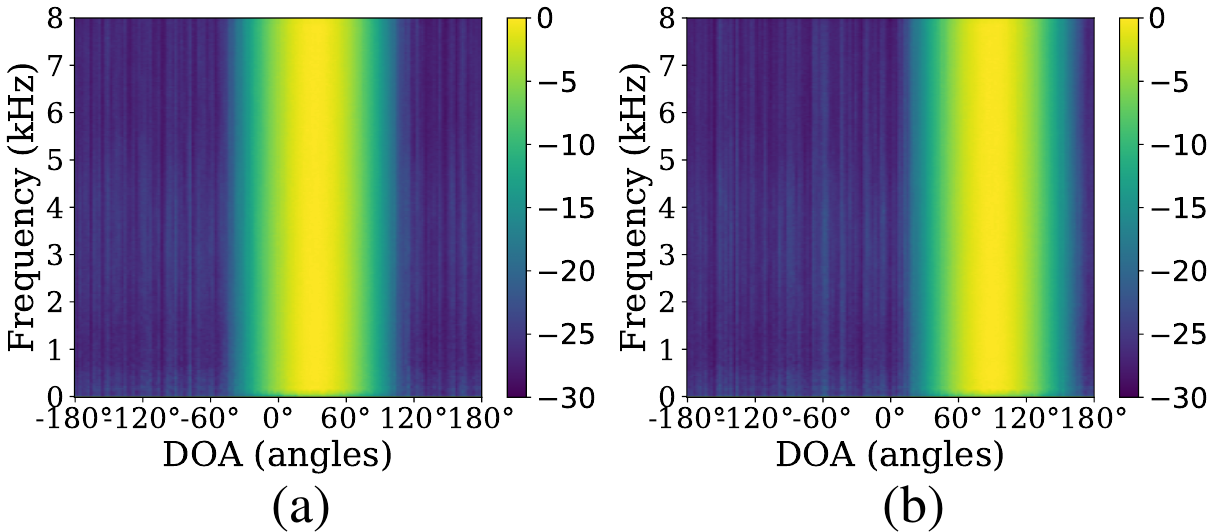} % double-cloumn: width=0.95
\caption{\chg{Estimated narrowband power patterns of the $6^{\textrm{th}}$-order steerable \ac{NDF} model. (a) steering direction $\theta_\textrm{s}=32.5^{\circ}$; (b) $\theta_\textrm{s}=90^{\circ}$.}}
\label{fig:sdr_ssf_pattern}
\end{figure}

We trained three \ac{NDF} models with the steering mechanism for the $1^{\textrm{st}}$-, $3^{\textrm{rd}}$-, and $6^{\textrm{th}}$-order patterns. Table~\ref{tab:sdr_ssf} shows the \ac{SDR}, SCOREQ, and \ac{PESQ} performance of the models with the mainlobe steered towards $\theta_\textrm{s} \in \{0^{\circ},30^{\circ},32.5^{\circ}, 60^{\circ},67.5^{\circ}, 90^{\circ}\}$. The narrowband power pattern estimates for the $6^{\textrm{th}}$-order pattern for $\theta_\textrm{s} \in \{32.5^{\circ}, 90^{\circ}\}$  are shown in Figure~\ref{fig:sdr_ssf_pattern}. We observe that the steerable \ac{NDF} models achieve similar performance across different steering directions and maintain frequency invariance, even though $32.5^{\circ}$ and $67.5^{\circ}$ were not included during training.

\subsection{Patterns with user-defined shapes}
\begin{figure}[t!]
\centering
\includegraphics[width=0.95\reprintcolumnwidth]{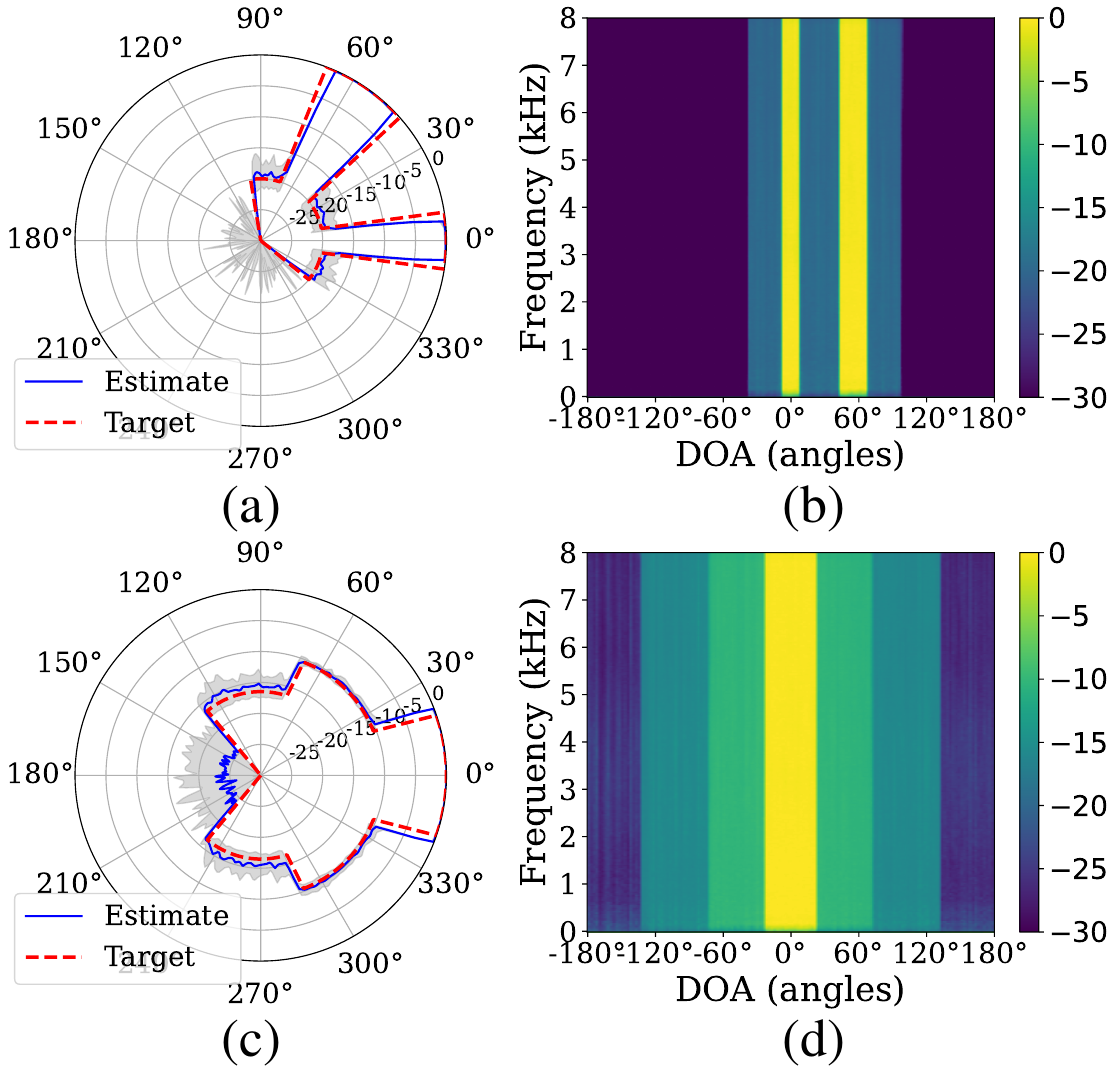} % double-cloumn: width=0.95
\caption{Analysis of two user-defined patterns. \rev{(a),(b) first user-defined pattern; (c),(d) second user-defined pattern; (a),(c) wideband $\widehat{\mathcal{P}}[\theta]$; (b),(d) narrowband $\widehat{\mathcal{P}}[\theta,f]$.} The gray area in polar plots represents the standard deviation of the estimate.}
\label{fig:manual_pattern}
\end{figure}

The shape of a specific pattern trained for \ac{NDF} is determined solely by the target \ac{VDM} signal, and the architecture or the loss function is not confined to one particular pattern. To illustrate this, we explore the learning of patterns with user-defined shapes. Figure~\ref{fig:manual_pattern} shows the target patterns we explored. The first pattern in Figures~\ref{fig:manual_pattern}(a) and (b) has two mainlobes of widths $20^\circ$ and $30^\circ$ with $0$~\unit{\decibel} attenuation, and a broad null region, while the second pattern in Figures~\ref{fig:manual_pattern}(c) and (d) has a step-like spatial pattern with sharp transitions in the attenuation levels and a null towards $180^\circ$. From Figure~\ref{fig:manual_pattern}, we observe that the unattenuated regions in both patterns are well approximated in a frequency-invariant manner, with a gradual transition at the boundaries. However, in Figure~\ref{fig:manual_pattern}(c), attenuation towards the null direction is limited to $-25$~\unit{\decibel}, and the variance of the estimated pattern increases for attenuation levels beyond $-15$~\unit{\decibel}, similar to observations for the DMA patterns.

\section{Evaluation in simulated reverberant environment}\label{sec:exp_rvb}
In this section, we present a comparative study of the \ac{NDF} models trained on anechoic and reverberant datasets, referred to as A-Model and R-Model, respectively. The study utilizes the directivity factor to measure the mask's impact on reverberant components, in conjunction with the power pattern and metrics reported in previous sections.

\chg{\subsection{Signal estimation quality}}

\begin{table}[t]
  \centering
  \caption{\rev{\chg{SDR (\si{\decibel}), SCOREQ, and PESQ on reverberant test sets.}}}
  \label{tab:sdr_results_rmodelvsamodel}
  \renewcommand{\arraystretch}{1.15}
  \begin{adjustbox}{max width=\columnwidth}
    \begin{tabular}{l l ccc @{\hskip 1.6em} ccc @{\hskip 1.6em} ccc}
      \toprule
      & &
      \multicolumn{3}{c}{$\mathrm{RT}_{60} = 0.2$~\unit{\s}} &
      \multicolumn{3}{c}{$\mathrm{RT}_{60} = 0.4$~\unit{\s}} &
      \multicolumn{3}{c}{$\mathrm{RT}_{60} = 0.6$~\unit{\s}} \\
      \cmidrule(lr){3-5} \cmidrule(lr){6-8} \cmidrule(lr){9-11}
      Method & Pattern & SDR & SCOREQ & PESQ & SDR & SCOREQ & PESQ & SDR & SCOREQ & PESQ \\
      \midrule
      DMA \citep{huang2020continuously} & $1^{\textrm{st}}$-order &
      6.82 & 0.97 & 2.48 & 7.71 & 0.80 & 2.77 & 7.92 & 0.71 & 2.88 \\
      LS Beamformer \citep{ls_beamforming} & $1^{\textrm{st}}$-order &
      10.83 & 1.13 & 2.30 & 11.62 & 0.98 & 2.67 & 11.78 & 0.87 & 2.86 \\
      \midrule
      \multirow{3}{*}{NDF A-Models}
        & $1^{\textrm{st}}$-order \ \ \ & 19.43 & 0.79 & 4.26 & 18.23 & 0.61 & 4.31 & 17.75 & 0.52 & 4.31 \\
        & $3^{\textrm{rd}}$-order \ \ \ & 11.81 & 0.72 & 3.84 &  9.27 & 0.60 & 3.83 &  8.59 & 0.52 & 3.83 \\
        & $6^{\textrm{th}}$-order \ \ \ &  8.34 & 0.68 & 3.54 &  5.64 & 0.61 & 3.35 &  4.90 & 0.53 & 3.31 \\
      \midrule
      \multirow{3}{*}{NDF R-Models}
        & $1^{\textrm{st}}$-order \ \ \ & 22.12 & 0.78 & 4.38 & 20.37 & 0.60 & 4.40 & 19.70 & 0.51 & 4.40 \\
        & $3^{\textrm{rd}}$-order \ \ \ & 14.30 & 0.69 & 4.06 & 11.59 & 0.56 & 4.05 & 10.74 & 0.49 & 4.03 \\
        & $6^{\textrm{th}}$-order \ \ \ & 10.58 & 0.65 & 3.79 &  7.77 & 0.55 & 3.65 &  6.92 & 0.48 & 3.59 \\
      \bottomrule
    \end{tabular}
  \end{adjustbox}
\end{table}

\chg{Table~\ref{tab:sdr_results_rmodelvsamodel} shows a comparison of the \ac{SDR}, SCOREQ, and \ac{PESQ} achieved by the R-Model, A-Model, and the baselines (DMA and \ac{LS} beamformer \citep{ls_beamforming}) for reverberation times 0.2~\unit{\s}, 0.4~\unit{\s}, and 0.6~\unit{\s}.} We observe that both the R-Model and A-Model outperform baselines under various reverberation conditions, underscoring the effectiveness of the \ac{NDF} models, and the R-Models consistently achieve better performance than the A-Models across different reverberation conditions, regardless of the order of the learned patterns. This illustrates the effectiveness of our training strategy for reverberant environments. However, \ac{SDR} and \ac{PESQ} performance decreases with both pattern order and reverberation time, with pattern order having a much larger effect.

\subsection{Power patterns}
\label{ssec:rev_power_patterns}
\begin{figure}[t!]
\centering
\includegraphics[width=0.95\reprintcolumnwidth]{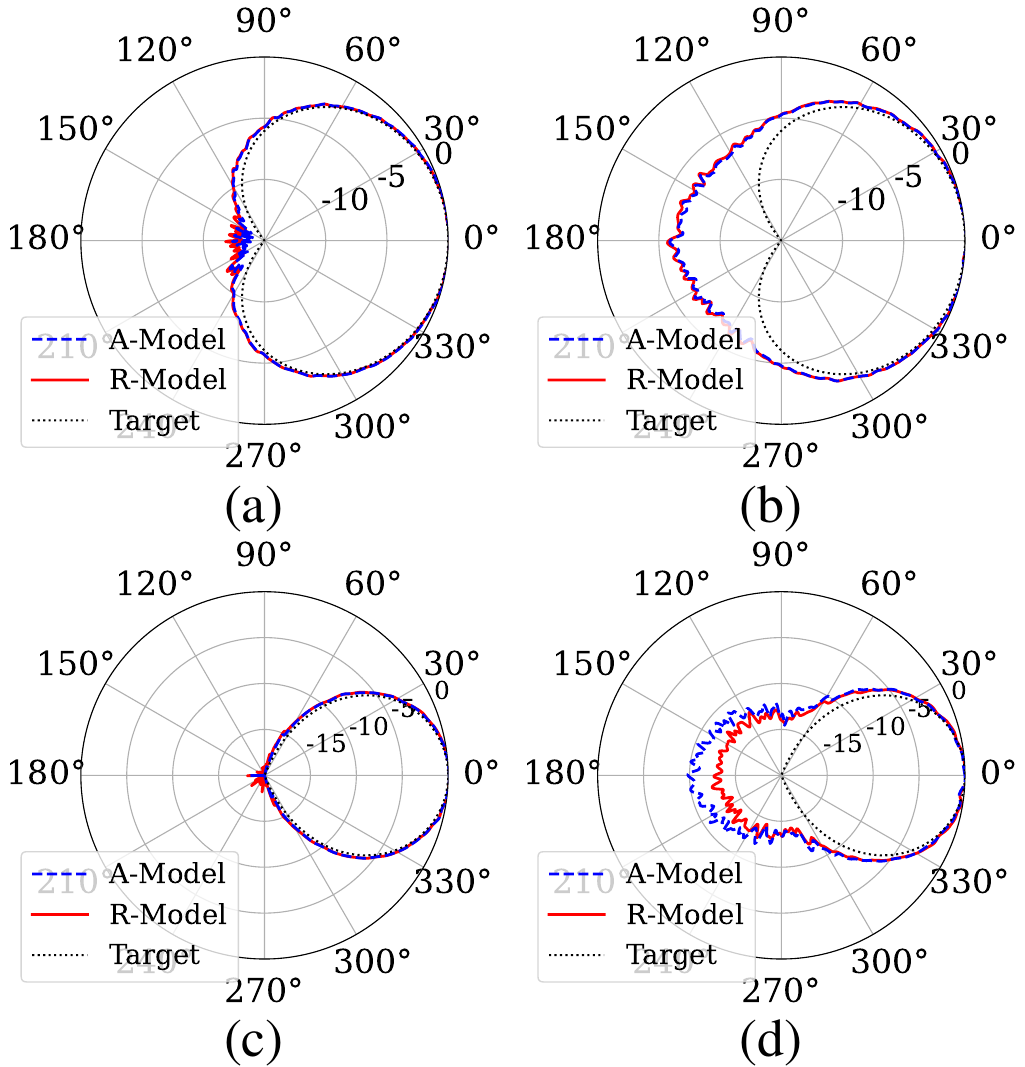} % double-cloumn: width=0.95
\caption{Comparison of the estimated wideband power patterns for the A-Model and R-Model. The source-array distances were fixed at 1~\unit{\metre}. \rev{The top and bottom rows correspond to the $1^{\textrm{st}}$-order and $6^{\textrm{th}}$-order patterns; the left and right columns correspond to $\textrm{RT}_{60}=0.2$~\unit{\s} and $\textrm{RT}_{60}=0.6$~\unit{\s}, respectively.}}
\label{fig:wb-bp-rvb}
\end{figure}

\begin{figure}[t!]
\centering
\includegraphics[width=0.95\reprintcolumnwidth]{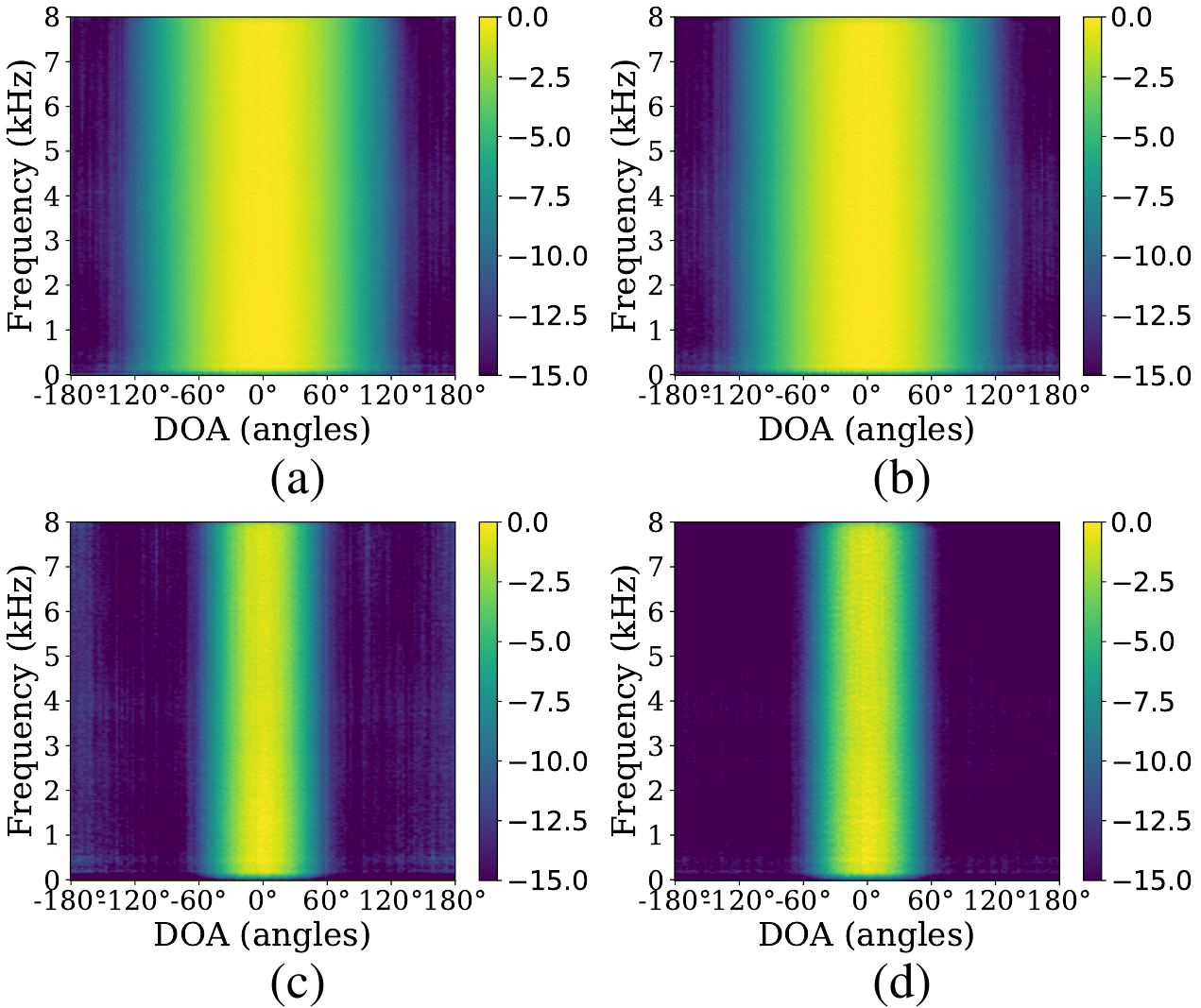} % double-cloumn: width=0.95
\caption{\chg{Comparison of the estimated narrowband patterns between the A-Model and R-Model. The left and right columns correspond to A-Model and R-Model, respectively. \rev{The \ac{NDF} models corresponding to (a) and (b) are trained for $1^{\textrm{st}}$-order pattern. The \ac{NDF} models corresponding to (c) and (d) are trained for $6^{\textrm{th}}$-order pattern.}}}
\label{fig:nb-bp-rvb}
\end{figure}

We further analyze the obtained power patterns of R-Models and A-Models in environments with low reverberation ($\textrm{RT}_{60} = 0.2$~\unit{\s}) and high reverberation ($\textrm{RT}_{60} = 0.6$~\unit{\s}). To effectively investigate the impact of masks on the direct-path components used to estimate the power patterns, we set the two concurrent sources in each test sample at a fixed source-array distance of 1~\unit{\m}. This distance results in a positive \ac{DRR}. Figure~\ref{fig:wb-bp-rvb} compares the estimated wideband power patterns for the A-Model and R-Model, respectively. For the $1^{\textrm{st}}$-order pattern, both models estimate the pattern similarly, suggesting that for lower-order patterns that are easier to learn, the A-Model, trained in non-reverberant conditions, can handle direct-path components as well as the R-Model. This phenomenon is also observed for a reverberation time of 0.2~\unit{\s}. However, for the $6^{\textrm{th}}$-order patterns in longer reverberation time (0.6~\unit{\s}), the R-Models demonstrate a higher suppression of direct-path sound in the region surrounding the null position than the A-Model. Figure~\ref{fig:nb-bp-rvb} shows the approximated narrowband power patterns for $1^{\textrm{st}}$-order under $\textrm{RT}_{60}= 0.2$~\unit{\s} and for $6^{\textrm{th}}$-order under $\textrm{RT}_{60}= 0.6$~\unit{\s}, which represent the easiest and most challenging setting, respectively. These results also show that the estimated power patterns in reverberant environments remain frequency-invariant.

\begin{figure}[t!]
\centering
\includegraphics[width=0.95\reprintcolumnwidth]{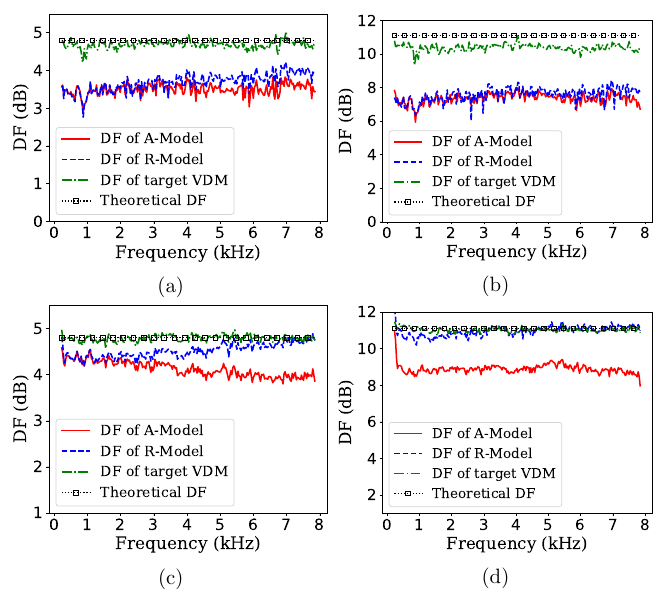} % double-cloumn: width=0.95
\caption{\chg{Estimated DF comparison between R-Model and A-Model. The source-array distance was fixed at 2.5~\unit{\metre}. \rev{(a) $1^{\textrm{st}}$-order, $\textrm{RT}_{60}=0.2$~\unit{\s}; (b) $6^{\textrm{th}}$-order, $\textrm{RT}_{60}=0.2$~\unit{\s}; (c) $1^{\textrm{st}}$-order, $\textrm{RT}_{60}=0.6$~\unit{\s}; (d) $6^{\textrm{th}}$-order, $\textrm{RT}_{60}=0.6$~\unit{\s}.}}}
\label{fig:DI_rvb}
\end{figure}

\subsection{Directivity factor}
We now focus on the \ac{DF} obtained by the \ac{NDF} models. As the \ac{DF} is computed based on the reverberant components, we set the source-array distance to 2.5~\unit{\m} (low DRR condition). Figure~\ref{fig:DI_rvb} shows the frequency-dependent \acp{DF} of the A-Models and R-Models trained for $1^{\textrm{st}}$-order and $6^{\textrm{th}}$-order patterns and evaluated in simulated rooms with a reverberation time of $0.2$ and $0.6$~\unit{\s}. We observe the following: Firstly, the R-Model mostly outperforms the A-Model in terms of \ac{DF}, particularly for higher reverberation time, which aligns with the signal quality results in Table~\ref{tab:sdr_results_rmodelvsamodel}. Secondly, as the reverberation time increases, the \ac{DF} of both models tends to increase; however, the R-Model exhibits a higher increase. Under the condition of $\textrm{RT}_{60} = 0.6~\unit{\s}$, the \ac{DF} of the R-Model tends to approach or even surpass the \ac{DF} of the \ac{VDM} target. This indicates that the R-Model tends to slightly over-suppress reverberation under $\textrm{RT}_{60} = 0.6~\unit{\s}$ that is more reverberant than the highest $\textrm{RT}_{60} = 0.5~\unit{\s}$ encountered during its training. This stronger suppression reflects genuine reverberation reduction rather than distortion of the target signal, as the power patterns in Sec.~\ref{ssec:rev_power_patterns} show a 0~\unit{\decibel} response along the steering direction, indicating that the direct sound from the target direction is preserved without attenuation. Since the \ac{DF} estimates are expected to be more accurate in higher reverberation conditions, this further demonstrates that the R-Model has better capabilities to handle reverberation. Thirdly, the \ac{DF} calculated for the target \ac{VDM} closely matches the theoretical \ac{DF} values of the various target patterns, particularly the $1^{\textrm{st}}$-order pattern or $\textrm{RT}_{60} = 0.6~\unit{\s}$. This demonstrates the accuracy of the proposed DF computation for the target \ac{VDM}.

\section{Applications with moving sources}\label{sec:unseen}
In the \ac{NDF} training strategy presented in Section~\ref{sec:pm}, the speech sources are assumed to be stationary during training. Therefore, the evaluation in Sections~\ref{sec:exp_setup}--\ref{sec:exp_rvb} focused on stationary source scenarios. In this section, we illustrate the performance of \ac{NDF} models trained using static sources in a moving source scenario. We consider two application scenarios: a mono audio recording in a simulated environment and a stereo audio recording in a real room. Audio examples can be found online\footnote{\url{https://www.audiolabs-erlangen.de/resources/2026-JASA-NDF}}.

\begin{figure}[t!]
\centering
\includegraphics[width=0.45\reprintcolumnwidth]{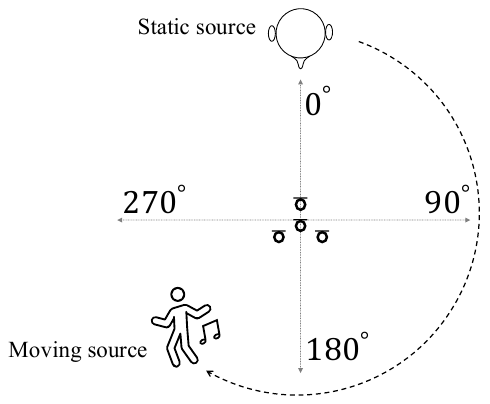} % double-cloumn: width=0.45
\caption{Simulated two-source scenario with a static speech source and a moving music source. The source at $0^{\circ}$ was static, while the moving source completed a full circle around the array in the clockwise direction. The source-array distance was $1.5$~\unit{\m}.}
\label{fig:interfering_moving_source}
\end{figure}

\begin{figure}[t!]
\centering
\includegraphics[width=0.95\reprintcolumnwidth]{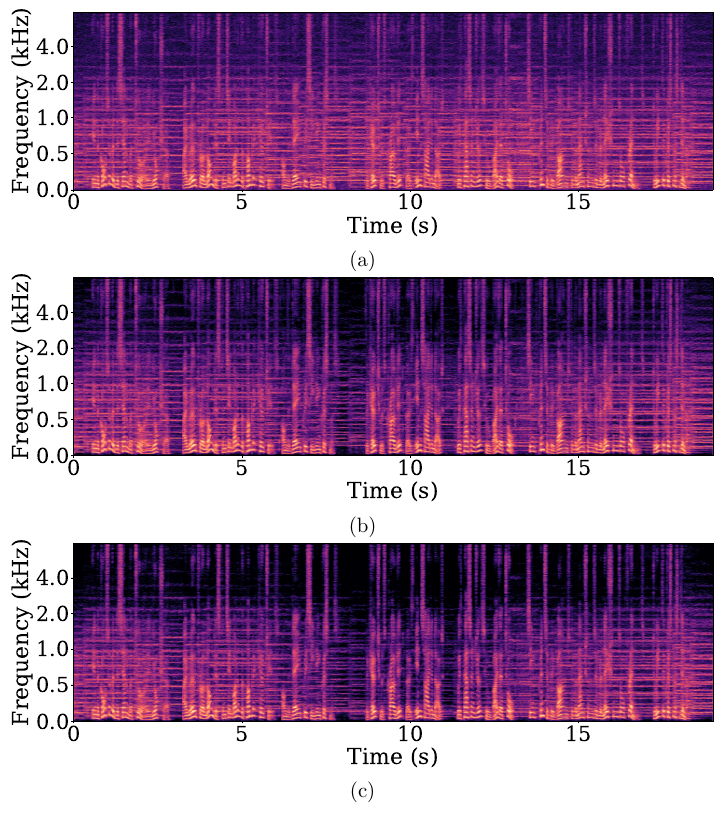} % double-cloumn: width=0.95
\caption{Comparison of spectrograms for a simulated moving scenario. \rev{(a) reference microphone signal; (b) target \ac{VDM} signal; (c) output signal of the \ac{NDF} model.}}
\label{fig:spectroCompare}
\end{figure}

\subsection{\chg{Mono audio recording}}
The acoustic scene and recording setup for this scenario are depicted in Figure~\ref{fig:interfering_moving_source}, which contains two sources: a stationary speech source and a moving music source, both co-planar with the microphone array, at a fixed distance of $1.5$~\unit{\m} from the array center. The stationary source was located at $0^{\circ}$, and the moving source completed one full rotation around the array in approximately $18$~\unit{\s} at a constant speed. The simulated room is 5~\unit{\m} $\times$ 4~\unit{\m} $\times$ 3.5~\unit{\m} and has an $\textrm{RT}_{60}$ of 0.15~\unit{\s}. The \ac{NDF} model, trained in anechoic environments with a static $1^{\textrm{st}}$-order Cardioid pattern pointing at $0^{\circ}$, is used for demonstration.

Figures~\ref{fig:spectroCompare}(a) and (b) show the spectrograms of the mixture signal at the reference microphone and the target \ac{VDM} signal, respectively. In the target \ac{VDM} signal, we observe that the amplitude of the music signal gradually decreases as it moves towards the null direction ($180^\circ$), followed by a gradual restoration to the original levels as it completes a full rotation, consistent with the desired spatial response. Figure~\ref{fig:spectroCompare}(c) shows the \ac{NDF} output, which follows the target \ac{VDM} signal, except for a slightly stronger suppression of the music source at higher frequencies and near the null direction. Notably, no visible distortion of the speech signal from the desired direction is observed in (b) or (c). 

\subsection{Stereo audio recording}
\begin{figure}[t!]
\centering
\includegraphics[width=0.55\reprintcolumnwidth]{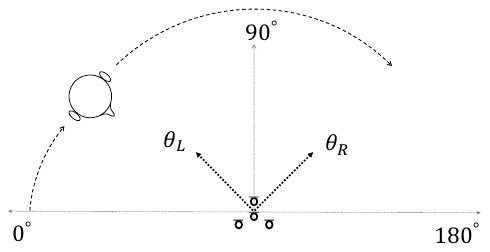} % double-cloumn: width=0.55
\caption{The scenario for stereo audio recording. One active speaker is moving from $0^{\circ}$ to $180^{\circ}$ with a fixed distance of $1.5$~m. $\theta_\textrm{L} = 45^{\circ}$ and $\theta_\textrm{R}= 135^{\circ}$ stand for two different steering directions of the power pattern. The power pattern learned by \ac{NDF} is $1^{\textrm{st}}$-order Cardioid pattern.}
\label{fig:waveformCompare_stereo_realrec_s1}
\end{figure}

A stereo audio recording can be made using two co-located $1^{\textrm{st}}$-order Cardioid microphones pointing to $45^{\circ}$ and $135^{\circ}$ \citep{williams2002stereophonic}. We investigate the application of the \ac{NDF} to perform a stereo audio recording. To this end, we enacted the acoustic scene in a real room ($4.6$~\unit{\m} $\times$ $4.5$~\unit{\m} $\times$ $2.6$~\unit{\m}) with $\textrm{RT}_{60}= 0.23$~\unit{\s}, depicted in Figure~\ref{fig:waveformCompare_stereo_realrec_s1}. \chg{As shown, the scene consisted of a real male speaker going from $0^\circ$ to $180^\circ$ in a clockwise direction, moving for approximately 8~\unit{\s} while maintaining an approximate distance of 1.5~\unit{\m} from the array center.} The recording was processed with the $1^{\textrm{st}}$-order Cardioid steerable NDF model steered towards $45^{\circ}$ and $135^{\circ}$, and the resulting audio outputs were assigned to the left and right channels of the stereo audio.

\begin{figure}[t!]
\centering
\includegraphics[width=0.65\reprintcolumnwidth]{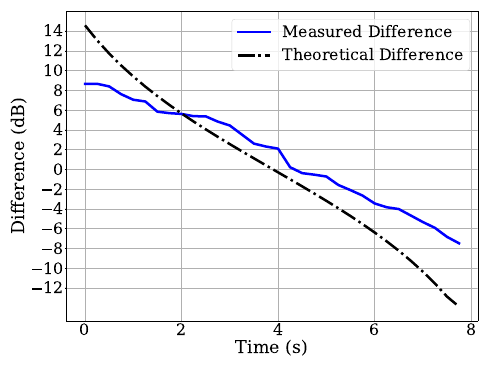} % double-cloumn: width=0.65
\caption{\chg{Amplitude difference between left channel and right channel in a real room with $\textrm{RT}_{60}= 0.23$~\unit{\s}.}}
\label{fig:DifferenceCompare_stereo_realrec}
\end{figure}

Figure~\ref{fig:DifferenceCompare_stereo_realrec} shows the segmental amplitude difference between the left and right channels, computed using segments of duration 1~\unit{\s} with a $75\%$ overlap between successive segments. We see that the level differences between the left and right channels of the stereo recording are effectively captured in the \ac{NDF} outputs, \chg{with a measured difference of 16~\unit{\decibel}}. However, there is still a gap compared to the theoretical value. Theoretically, a Cardioid pattern could exhibit strong suppression near the null positions, a capability that is not fully realized in practice.

\rev{\section{Discussion of limitations}\label{sec:limitations}
While the preceding sections demonstrate the capabilities of \ac{NDF}, we discuss the limited null depth and summarize the conditions under which its performance may degrade.

\emph{Null depth.} For the scenarios under test, Figures~\ref{fig:narrowband_bp} and \ref{fig:manual_pattern}(c) show that the achieved attenuation saturates at approximately $-25$~\unit{\decibel} for both \ac{DMA} and user-defined patterns, and that the variance of the estimated pattern increases for attenuation levels beyond $-15$~\unit{\decibel}. We hypothesize that this is caused by the training process when sources lie both in the mainlobe direction and close to the null. Since the loss compares the estimated and the target \ac{VDM} signal as a whole, it is dominated by the mainlobe source: once the mainlobe components are estimated accurately, the overall signal is already close to the target, and the loss is small, largely regardless of how accurately the null sources are estimated. In contrast, an error in the mainlobe changes the signal substantially and is immediately reflected in the loss. Moreover, during directivity-pattern estimation, even a small absolute error for sources near the null can appear as a large deviation in decibels. This may account for both the saturation of null depth and the increased variance at high attenuation levels. Further improving the directional response, especially in the null directions, remains a topic for future work.

\emph{Adverse conditions.} The experimental results identify three conditions under which the estimated pattern and \ac{VDM} signal deviate from the target. (i) At a very low sensor \ac{SNR} on a compact array (Figure~\ref{fig:bp_diameters_snr}(b)), the \ac{NDF} renders an omnidirectional response below approximately 250~\unit{\hertz} with a gain above 0~\unit{\decibel}, analogous to the white-noise amplification of differential beamformers. %This essentially results in a natural deviation of the target directivity pattern. 
(ii) Above the spatial aliasing frequency, the pattern is only preserved when sufficient information below the aliasing frequency is available (Figures~\ref{fig:bp_plots}(c) and (d)). \ac{NDF} applied to high-frequency narrowband signals would therefore suffer from spatial aliasing. (iii) The signal estimation quality degrades with increasing pattern order and reverberation time, as shown in Table~\ref{tab:sdr_results_rmodelvsamodel}.
}

\section{Conclusions}\label{sec:cls}
Neural directional filtering (NDF) offers a viable solution for a challenging task: capturing sound with a controllable directivity pattern using a compact microphone array. In this paper, we introduced an effective training strategy that enables the NDF model to learn different patterns and enhances its ability to operate in reverberant environments. We analyzed the performance of \ac{NDF} on both direct-path components and reverberant components of reverberant signals, utilizing estimated directivity patterns and directivity factors. Additionally, we conducted a comprehensive study on the processing mechanisms and characteristics, including its pattern-learning capabilities (such as maintaining frequency-invariant patterns, mitigating spatial aliasing, learning higher-order DMA patterns, and user-defined patterns), as well as its applications to moving sources.

\begin{acknowledgments}
The authors gratefully acknowledge the scientific support and HPC resources provided by the Erlangen National High Performance Computing Center (NHR@FAU) of the Friedrich-Alexander-Universit\"{a}t Erlangen-N\"{u}rnberg (FAU). The hardware is funded by the German Research Foundation (DFG). The authors thank Mr. Julian Wechsler for his contributions to the initial work.
\end{acknowledgments}

%%%%% JASA-required sections %%%%%

\section*{AUTHOR DECLARATIONS}
\subsection*{Conflict of Interest}
The authors have no conflicts to disclose.

\section*{DATA AVAILABILITY}
The data that support the findings of this study are available from the corresponding author upon reasonable request.

%%%%% Bibliography %%%%%
\bibliographystyle{jasanum2}
\bibliography{refs}

\end{document}